\documentclass[twocolumn,english,prb,aps,english,superscriptaddress]{revtex4}
\usepackage[T1]{fontenc}
\usepackage[latin9]{inputenc}
\usepackage{color}
\usepackage{float}
\usepackage{amsmath}
\usepackage{amssymb}
\usepackage{graphicx}
\usepackage{esint}

\makeatletter

\@ifundefined{textcolor}{}
{%
 \definecolor{BLACK}{gray}{0}
 \definecolor{WHITE}{gray}{1}
 \definecolor{RED}{rgb}{1,0,0}
 \definecolor{GREEN}{rgb}{0,1,0}
 \definecolor{BLUE}{rgb}{0,0,1}
 \definecolor{CYAN}{cmyk}{1,0,0,0}
 \definecolor{MAGENTA}{cmyk}{0,1,0,0}
 \definecolor{YELLOW}{cmyk}{0,0,1,0}
}

\@ifundefined{textcolor}{}{%
 \definecolor{BLACK}{gray}{0}
 \definecolor{WHITE}{gray}{1}
 \definecolor{RED}{rgb}{1,0,0}
 \definecolor{GREEN}{rgb}{0,1,0}
 \definecolor{BLUE}{rgb}{0,0,1}
 \definecolor{CYAN}{cmyk}{1,0,0,0}
 \definecolor{MAGENTA}{cmyk}{0,1,0,0}
 \definecolor{YELLOW}{cmyk}{0,0,1,0}
}

\usepackage{dcolumn}
\usepackage{bm}
\usepackage{enumerate}

\usepackage{babel}

\usepackage{babel}

\makeatother

\usepackage{babel}
\begin{document}

\title{Nonequilibrium quantum systems with electron-phonon interactions:
Transient dynamics and approach to steady state}

\author{Eli Y. Wilner}

\affiliation{School of Physics and Astronomy, The Sackler Faculty of Exact Sciences,
Tel Aviv University,Tel Aviv 69978,Israel}

\author{Haobin Wang}

\affiliation{Department of Chemistry and Biochemistry, New Mexico State University,
Las Cruces, NM 88003, USA}

\author{Michael Thoss}

\affiliation{Institute for Theoretical Physics and Interdisciplinary Center for
Molecular Materials, Friedrich-Alexander-Universität Erlangen-Nürnberg,
Staudtstr. 7/B2, 91058 Erlangen, Germany}

\author{Eran Rabani}

\affiliation{School of Chemistry, The Sackler Faculty of Exact Sciences, Tel Aviv
University,Tel Aviv 69978,Israel}
\begin{abstract}
The nonequilibrium dynamics of a quantum dot with electron-phonon
interactions described by a generalized Holstein model is presented.
A combination of methodologies including the reduced density matrix
formalism, the multilayer multiconfiguration time-dependent Hartree
method, and a time-dependent nonequilibrium Green function approach,
is used to explore the transient behavior on multiple timescales as
the system approaches steady-state. The dot population dynamics on
short to intermediate times is governed by the dot-lead hybridization
parameter ($\Gamma$) and by the typical phonon frequency ($\omega_{c}$)
and depends on the location of the energy level of the dot relative
to the bias window. At longer times, the dynamics show a distinct
behavior depending on whether the system is in the adiabatic or non-adiabatic
regime, with a quantum dot occupation that may depend on the initial
preparation of the phonons degrees of freedom. A ``phase'' diagram
of this localization effect as a function of the polaron shift ($\lambda$)
for various phonon frequencies is derived, suggesting the existence
of bistability on experimentally observable timescales.
\end{abstract}
\maketitle

\section{Introduction\label{sec:Introduction}}

The study and understanding of nonequilibrium phenomena in many-body
quantum systems has been of great interest recently. Among the variety
of architectures and processes considered, energy and charge transport
in nanostructures such as, e.g., single molecule junctions, carbon
nanotubes, and small quantum dots have received particular attention.\cite{Feldman08,Cuevas2010}
In contrast to mesoscopic or bulk systems, these nanosystems often
exhibit strong electron-phonon/vibrational interactions, which manifests
itself in interesting transport phenomena.\cite{galperin_hysteresis_2005,Leturcq09,Granger12,Benyamini14}
In molecular junctions, for example, electron-phonon interaction has
been shown to result in a multitude of nonequilibrium phenomena such
as current-induced local heating and cooling, multistability, switching
and hysteresis, as well as decoherence.\cite{galperin_hysteresis_2005,Galperin07,Leon08,Ioffe08,Natelson08,Haertle09,Huettel09,Repp10,Tao10,Ballmann10,Osorio10,Arroyo10,Secker11,Haertle11b,Kim11,Ward11,Ballmann12,Albrecht12,Ballmann13,wilner_bistability_2013,Haertle13,Haertle13b}

Most of the studies so far have focused on phenomena in steady-state.
Much less is known about transient dynamics in nanostructures under
nonequilibrium conditions. Fundamental questions to be addressed include:
What are the timescales on which a steady-state is reached under nonequilibrium
conditions? Which dynamical processes are of importance? What are
the underlying relaxation mechanisms? What are the preconditions for
the existence of a unique steady-state? In fact, the existence of
a unique steady-state in many-body quantum systems with electron-phonon
interaction has been a topic of great controversy in recent years.\cite{galperin_hysteresis_2005,Mitra2005,galperin_non-linear_2008,Bratkovsky2007,Bratkovsky2009,Kosov2011,albrecht_bistability_2012,Komnik2013,Albrecht2013,wilner_bistability_2013}

In this paper, we address these questions for a generic model of charge
transport in a quantum dot with electron-phonon interaction using
a reduced density matrix (RDM) formalism based on projection-operator
techniques.\cite{Leijnse2008,cohen_memory_2011,cohen_generalized_2013,wilner_bistability_2013}
This formalism requires as input the memory kernel. To this end, we
employ two different approaches: (i) a two-time nonequilibrium Green
function (NEGF) method and (ii) the multilayer multiconfiguration
time-dependent Hartree (ML-MCTDH)~\cite{Thoss03,Wang2009} approach.
The latter approach provides a numerically exact treatment of the
nonequilibrium dynamics within a certain timescale. Because the memory
kernel decays typically on a much shorter timescale than the RDM matrix
itself, this strategy allows a significant extension of the timescale
accessible by numerically exact ML-MCTDH technique and by the two-time
NEGF approach. This was demonstrated already in previous studies of
impurity models with electron-electron~\cite{cohen_generalized_2013,cohen_memory_2011}
and electron-phonon~\cite{wilner_bistability_2013} interactions. 

It should be noted that a variety of other approaches have been developed
and applied to study transient phenomena in nonequilibrium quantum
systems with electron-phonon interaction, including approximate methods
such as master equation methods,\cite{Kohler05,Li_Schreiber07,Li_Fainberg10}
as well as numerically exact schemes, such as numerical path-integral
approaches~\cite{muhlbacher_real-time_2008,Huetzen12,Simine13} and
the scattering state numerical renormalization group technique.\cite{Jovchev13}
The approaches employed in the present work allow a significant extension
of such studies with respect to the complexity of the phonon bath,
the range of physical parameters, and the accessible timescales.

The remainder of the paper is organized as follows. The model and
the theoretical methodology is outlined in Sec.\ \ref{sec:Model-and-Theoretical}.
In Sec.\ \ref{sec:Time-Scales-Analysis}, we analyze the quantum
dynamics, in particular with respect to the different timescales inherent
in the transient dynamics and the approach to steady-state. The dependence
of the dynamics on the initial preparation is discussed in Sec.\ \ref{sec:bistability}.
Sec.\ \ref{sec:Conclusions} concludes with a summary.

\section{Model and Theoretical Framework\label{sec:Model-and-Theoretical}}

\subsection{Model Hamiltonian}

We consider a generic model for charge transport through a quantum
dot with electron-phonon interaction, often referred to as the extended
nonequilibrium Holstein model. The model is described by the Hamiltonian:
\begin{equation}
H=H_{S}+H_{B}+V_{SB}\label{eq:Hamiltonian}
\end{equation}
where

\begin{equation}
H_{S}=\varepsilon_{d}d^{\dagger}d\label{eq:Hamiltonian-system}
\end{equation}
is the system (quantum dot) Hamiltonian, comprising a single electronic
state with energy $\varepsilon_{d}$ and corresponding fermionic creation/annihilation
operators $d^{\dagger}$/$d$. The bath is described by the sum of
fermionic leads and bosonic modes, $H_{B}=H_{\ell}+H_{{\rm ph}}$,
where

\begin{equation}
H_{\ell}=\sum_{k\in L,R}\varepsilon_{k}a_{k}^{\dagger}a_{k}\label{eq:Hamiltonian-leads}
\end{equation}
represents the noninteracting left/right ($L/R$) leads Hamiltonian
with fermionic creation/annihilation operators $a_{k}^{\dagger}$/$a_{k}$.
The bosonic bath Hamiltonian representing the phonons is given by:

\begin{equation}
H_{{\rm ph}}=\sum_{\alpha}\hbar\omega_{\alpha}\left(b_{\alpha}^{\dagger}b_{\alpha}+\frac{1}{2}\right)\label{eq:Hamiltonian-phonons}
\end{equation}
where $b_{\alpha}^{\dagger}$/$b_{\alpha}$ are the ladder operators
for the phonon mode $\alpha$ with energy $\hbar\omega_{\alpha}$.
Finally, the coupling between the system and the baths is given by
\begin{equation}
V_{SB}=\sum_{k\in L,R}\left(t_{k}da_{k}^{\dagger}+t_{k}^{*}a_{k}d^{\dagger}\right)+d^{\dagger}d\sum_{\alpha}M_{\alpha}\left(b_{\alpha}^{\dagger}+b_{\alpha}\right)\label{eq:V}
\end{equation}
where $t_{k}$ is the coupling strength between the system and lead
state $k$, determined from the relation 
\begin{equation}
\Gamma_{L,R}(\varepsilon)=2\pi\sum_{k\in L,R}|t_{k}|^{2}\delta(\varepsilon-\varepsilon_{k}).\label{eq:Gamma}
\end{equation}
Here, $\Gamma_{L,R}\left(\varepsilon\right)=\frac{a^{2}}{b^{2}}\sqrt{4b^{2}-(\varepsilon-\mu_{L,R})^{2}}$
is the electron spectral density, which is assumed to be of tight-binding
form, and $\mu_{L,R}$ is the chemical potential of the left/right
lead, respectively. In the applications reported below we choose typical
parameters for a metal lead, namely, $a=0.2\mbox{eV}$ and $b=1\mbox{eV}$.
For convenience the results in this paper are mostly presented in
dimensionless units scaled by $\Gamma$, where $\Gamma=0.16\mbox{eV}$
is the maximum value of $\Gamma_{R}\left(\varepsilon\right)+\Gamma_{L}\left(\varepsilon\right)$.

The second term in Eq.~(\ref{eq:V}) represents the electron-phonon
coupling, where $M_{\alpha}$ is the coupling strength to mode $\alpha$
determined from the relation 
\begin{equation}
J(\omega)=\pi\sum_{\alpha}M_{\alpha}^{2}\delta(\hbar\omega-\hbar\omega_{\alpha})
\end{equation}
where $J\left(\omega\right)=\frac{\pi\hbar}{2}\eta\omega e^{-\frac{\omega}{\omega_{c}}}$
is the phonon spectral function assumed to be of Ohmic form. The dimensionless
Kondo parameter, $\eta=\frac{2\lambda}{\hbar\omega_{c}}$, determines
the overall strength of the electron-phonon couplings, where $\omega_{c}$
is the characteristic phonon bath frequency and $\lambda=\sum_{\alpha}\frac{M_{\alpha}^{2}}{\hbar\omega_{\alpha}}=\frac{1}{\pi}\int\frac{d\omega}{\omega}J(\omega)$
is the reorganization energy (or polaron shift), which also determines
the shifting of the dot energy upon charging. We set realistic relaxation
timescales for the phonon bath, \textit{i.e.}, by choosing its characteristic
frequency, $\omega_{c}$, in the range of $25-1000\mbox{cm\ensuremath{^{-1}}}$,
which is $\approx0.02-0.8$ in units of $\Gamma/\hbar$.

The model introduced above and variants thereof have been widely used
to study nonequilibrium charge transport in nanostructures, such as,
for example, semiconductor quantum dots,\cite{Kubala10} carbon nanotubes~\cite{Leturcq09}
or molecular junctions.\cite{Cizek04,Mitra2004,Koch05,galperin_hysteresis_2005,Galperin07,Benesch08}
In the latter case, the phonons may include, in addition to the phonons
of the contacts, the vibrational degrees of freedom of the molecule.

\begin{figure}[t]
\includegraphics[width=8cm]{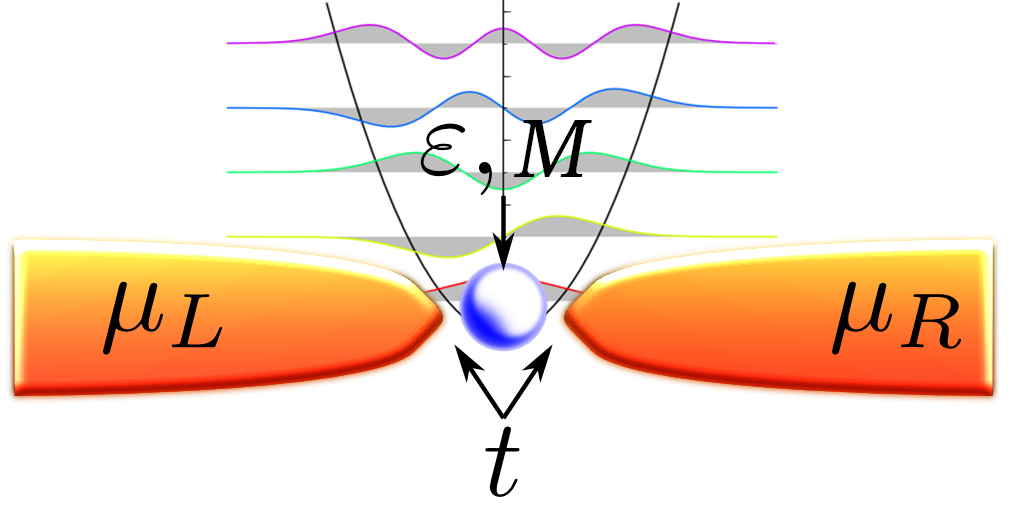} \caption{A sketch of the quantum dot coupled to left and right leads and to
a phonon bath.}

\label{fig:model} 
\end{figure}

\subsection{Reduced density matrix formalism}

To study the dynamic response on multiple timescales generated by
the extended Holstein model as the system is driven away from equilibrium,
we adopt the reduced density matrix (RDM) formalism~\cite{cohen_generalized_2013}
discussed in detail in Ref.\ \onlinecite{cohen_memory_2011}) for
the Anderson impurity model and in Ref.\ \onlinecite{wilner_bistability_2013}
for the present model. The equation of motion for the RDM, $\sigma(t)=Tr_{B}\{\rho(t)\}$,
is given by 
\begin{equation}
i\hbar\frac{\partial}{\partial t}\sigma\left(t\right)=\mathcal{L}_{S}\sigma\left(t\right)+\vartheta\left(t\right)-\frac{i}{\hbar}\int_{0}^{t}d\tau\kappa\left(\tau\right)\sigma\left(t-\tau\right)\label{eq:sigma(t)}
\end{equation}
where $\mathcal{L}_{S}=[H_{S},\cdots]$ is the system's Liouvillian,
$Tr_{B}\{\cdots\}$ is a trace over the baths degrees of freedom (leads
and phonon baths) and $\rho(t)$ is the full density matrix which
obeys the von-Neuman equation of motion. In the above, 
\begin{equation}
\vartheta\left(t\right)=Tr_{B}\left\{ \mathcal{L}_{V}e^{-\frac{i}{\hbar}Q\mathcal{L}t}Q\rho\left(0\right)\right\} \label{eq:theta(t)}
\end{equation}
depends on the choice of initial conditions and $\mathcal{L}_{v}=[V_{SB},\cdots]$.
By construction, $\vartheta\left(t\right)$ vanishes for an uncorrelated
initial state, \textit{i.e. }when $\rho(0)=\sigma(0)\otimes\rho_{B}(0)$,
where $\sigma(0)$ and $\rho_{B}(0)$ are the system and baths initial
density matrices, respectively. In all applications reported below
we start from a factorized initial condition and thus, ignore $\vartheta\left(t\right)$.
The memory kernel, which describes the non-Markovian dependency of
the time propagation of the system, is given by 
\begin{equation}
\kappa\left(t\right)=Tr_{B}\left\{ \mathcal{L}_{V}e^{-\frac{i}{\hbar}Q\mathcal{L}t}Q\mathcal{L}\rho_{B}\right\} \label{eq:memory-kernel}
\end{equation}
where $Q=1-P$, $P=\rho_{B}(0)Tr_{B}\{\cdots\}$ is a projection operator,
and $\mathcal{L}=[H,\cdots]$ is the full Liouvillian super-operator.

To obtain $\sigma(t)$, one requires as input the super-matrix of
the memory kernel. For a general system the super-matrix has $N^{4}$
elements, where $N$ is the dimension of the density matrix. Thus,
calculating all elements can be a tedious task.\cite{cohen_memory_2011}
The complexity is considerably reduced for the extended Holstein model.
First, $N=2$ and hence the memory kernel has only $16$ terms. Second,
the reduced dynamics of the diagonal elements of $\sigma(t)$ (the
populations) are decoupled from those of the off-diagonal elements
(the coherences). If one is interested in the populations alone (as
is the case in the present study), only $4$ elements of the memory
kernel are necessary to describe the population dynamics. To further
simplify the calculations of the memory, we express it in terms of
a Volterra equation of the second type, removing the complexity of
the projected dynamics of Eq.~(\ref{eq:memory-kernel}): 
\begin{equation}
\kappa\left(t\right)=i\hbar\dot{\Phi}\left(t\right)-\Phi\left(t\right)\mathcal{L}_{S}+\frac{i}{\hbar}\int_{0}^{t}d\tau\Phi\left(t-\tau\right)\kappa\left(\tau\right)\label{eq:volterra}
\end{equation}
with 
\begin{equation}
\Phi\left(t\right)=Tr_{B}\left\{ \mathcal{L}_{V}e^{-\frac{i}{\hbar}\mathcal{L}t}\rho_{B}\right\} .\label{eq:Phi(t)}
\end{equation}
Since the operator $\mathcal{L}_{V}$ appearing in the equation for
$\Phi(t)$ and the full Hamiltonian conserve the total particle number,
only the diagonal matrix elements $\Phi(t)$ need to be computed:
\begin{equation}
\Phi_{nn,mm}(t)=\frac{2}{\hbar}Tr_{B}\left\{ \rho_{B}\left\langle m\right|\sum_{k}t_{k}d(t)a_{k}^{\dagger}(t)\left|m\right\rangle \right\} ..\label{eq:phi(t)}
\end{equation}
Here, $|m\rangle$ denotes the electronic state of the quantum dot,
where $m$ can take the values $1$ or $0$, corresponding to an occupied
or an unoccupied dot, respectively. Note that $\Phi_{nn,mm}(t)$ is
independent on $n$ and thus has only 2 independent components. The
above expression for $\Phi_{nn,mm}(t)$ has a simple physical interpretation
as the time derivative of the dot population and can be expressed
in terms of the sum of the left ($I_{m}^{L}(t)$) and right ($I_{m}^{R}(t)$)
currents:

\begin{equation}
e\Phi_{nn,mm}(t)=I_{m}^{L}(t)+I_{m}^{R}(t),\label{eq:current-1}
\end{equation}
where 
\begin{equation}
I_{m}^{L,R}(t)=-\frac{2e}{\hbar}\Im\sum_{k\in L,R}t_{k}\langle m|d(t)a_{k}^{\dagger}(t)|m\rangle,\label{eq:current}
\end{equation}
is the left/right current for an initial occupied ($m=1$) or empty
($m=0$) dot, and $e$ is the electron charge.

\subsection{Calculation of the Memory Kernel}

The RDM formalism may seem redundant, since in order to obtain the
reduced density matrix one requires as input the memory kernel which
is given in terms of the left and right currents. If the left and
right currents are accessible by impurity solvers, so are the elements
of the RDM. This, however, ignores the fact that the memory kernel
typically decays on a much faster timescale compared to the RDM itself.\cite{cohen_memory_2011,cohen_generalized_2013,wilner_bistability_2013}
Thus, if the memory decays to zero at $t>t_{c}$ where $t_{c}$ is
a cutoff time, it is sufficient to obtain the memory kernel to $t_{c}$
and infer from that the dynamics of the RDM at all times. We refer
to this as the ``cutoff approximation'', which will become exact
if the memory kernel has a finite range and decays to zero at $t>t_{c}$.
Since numerical solvers of quantum impurity models scale exponentially
with the propagation time, this saves significant computational time.
As will be shown below, the RDM formalism provides means to study
the dynamics on timescales not accessible by direct impurity solvers.\cite{wilner_bistability_2013,Cohen2013kondo}

We adopt two impurity solvers to calculate the memory kernel. The
first is based on the so-called multilayer multiconfiguration time-dependent
Hartree theory in second quantization representation (ML-MCTDH-SQR)~\cite{Wang2009}
and the second, described below, is based on a two-time nonequilibrium
Green function (NEGF) formalism.

\subsubsection{Multilayer multiconfiguration time-dependent Hartree (ML-MCTDH) theory}

The ML-MCTDH theory is a rigorous variational method used for propagating
wave packets in complex systems with many degrees of freedom.\cite{Thoss03}
Extending the original MCTDH method,\cite{Meyer90,Meyer09} employs
a hierarchical, multilayer representation of the many-body wave function.
Originally developed for treating distinguishable particles, it has
recently been generalized to describe indistinguishable fermionic
or bosonic particles employing occupation number representation of
the Fock space in the second quantized framework.\cite{Wang2009}
The approach has been applied to nonequilibrium transport with electron-phonon~\cite{Wang2009,wang_numerically_2011,albrecht_bistability_2012,Wang13b}
and electron-electron interactions.\cite{Wang13} For completeness,
we provide a brief summary of this approach and its specific implementation
for calculating the memory kernel in the extended Holstein model. 

Within the ML-MCTDH, the wave function is represented by a recursive,
layered expansion 
\begin{equation}
\left|\Psi\left(t\right)\right\rangle =\sum_{j_{1}}\sum_{j_{2}}\cdots\sum_{j_{p}}A_{j_{1}j_{2}...j_{p}}\left(t\right)\prod_{\kappa=1}^{p}\left|\varphi_{j_{\kappa}}^{\left(\kappa\right)}\left(t\right)\right\rangle \label{eq:mcdth_1}
\end{equation}
\begin{equation}
\left|\varphi_{j_{\kappa}}^{\left(\kappa\right)}\left(t\right)\right\rangle =\sum_{i_{1}}\sum_{i_{2}}\cdots\sum_{i_{Q(\kappa)}}B_{i_{1}i_{2}...i_{Q\left(\kappa\right)}}^{\kappa,j_{k}}\left(t\right)\prod_{q=1}^{Q(\kappa)}\left|v_{i_{q}}^{\left(\kappa,q\right)}\left(t\right)\right\rangle \label{eq:mcdth_2}
\end{equation}
\begin{align}
\left|v_{i_{q}}^{\left(\kappa,q\right)}\left(t\right)\right\rangle  & =\sum_{\alpha_{1}}\sum_{\alpha_{2}}\cdots\sum_{\alpha_{M(\kappa,q)}}\nonumber \\
 & C_{\alpha_{1}\alpha_{2}...\alpha_{M\left(\kappa,q\right)}}^{\kappa,j_{k},i_{q}}\left(t\right)\prod_{q=1}^{M(\kappa,q)}\left|\xi_{\alpha_{\gamma}}^{\left(\kappa,q,\gamma\right)}\left(t\right)\right\rangle 
\end{align}
where $A_{j_{1}j_{2}...j_{p}},B_{i_{1}i_{2}...i_{Q\left(\kappa\right)}}^{\kappa,j_{k}},C_{\alpha_{1}\alpha_{2}...\alpha_{M\left(\kappa,q\right)}}^{\kappa,j_{k},i_{q}}$
and so on are the expansion coefficients for the first, second, third,
... , layers, respectively. $\left|\varphi_{j_{\kappa}}^{\left(\kappa\right)}\left(t\right)\right\rangle ,\left|v_{i_{q}}^{\left(\kappa,q\right)}\left(t\right)\right\rangle ,\left|\xi_{\alpha_{\gamma}}^{\left(\kappa,q,\gamma\right)}\left(t\right)\right\rangle $,...,
are the single particle functions for the first, second, third, ...
, layers. For distinguishable particles, the primitive basis functions
for each degree of freedom in the deepest layer can be any convenient
choice depending on the specific form of the Hamiltonian operator,
e.g., Fourier grid points, harmonic oscillator eigenfunctions, Legendre
polynomials, etc. When treating identical particles, a second quantization
representation (SQR) is employed, where the primitive basis functions
for each single particle group in the deepest layer are the occupation
number states of this Fock subspace.\cite{Wang2009} This is referred
to as the ML-MCTDH-SQR approach. In principle, the recursive multilayer
expansion/hierarchical tensor decomposition can be carried out to
an arbitrary number of layers. In practice, the multilayer hierarchy
is terminated at a particular level by expanding the single particle
functions in the deepest layer in terms of time-independent configurations/primitive
basis functions. The ML-MCTDH equations of motion are obtained by
applying Dirac-Frenkel variational principle to Eq.~(\ref{eq:mcdth_1}).\cite{Thoss03,Wang2009}
In the applications reported below, four dynamical layers are used
to represent the wave function.

Within a certain timescale, the electronic and phonon continua can
be discretized to and represented by a finite number of electronic
states and phonon modes. For the parameter regimes discussed in this
paper, a typical number of $300-400$ electronic states and $800-1200$
phonon modes were sufficient to achieve convergence (to within a few
percent relative error). Systematic test-calculations were then carried
out to check against the number of primitive basis functions and the
number of configurations for each layer until convergence was achieved.\cite{Thoss03,Wang2009}
The computed time-dependent multilayer wave functions were then used
to obtain the left and right currents $I_{0}^{L}(t)$, $I_{1}^{L}(t)$,
$I_{0}^{R}(t)$, and $I_{1}^{R}(t)$ and the currents were used to
generated the elements of $\Phi_{nn,mm}\left(t\right)$ and the corresponding
elements of the memory kernel were obtained by solving the Volterra
equation (cf., Eq.~\ref{eq:volterra}).

\begin{figure}[H]
\includegraphics[width=8cm]{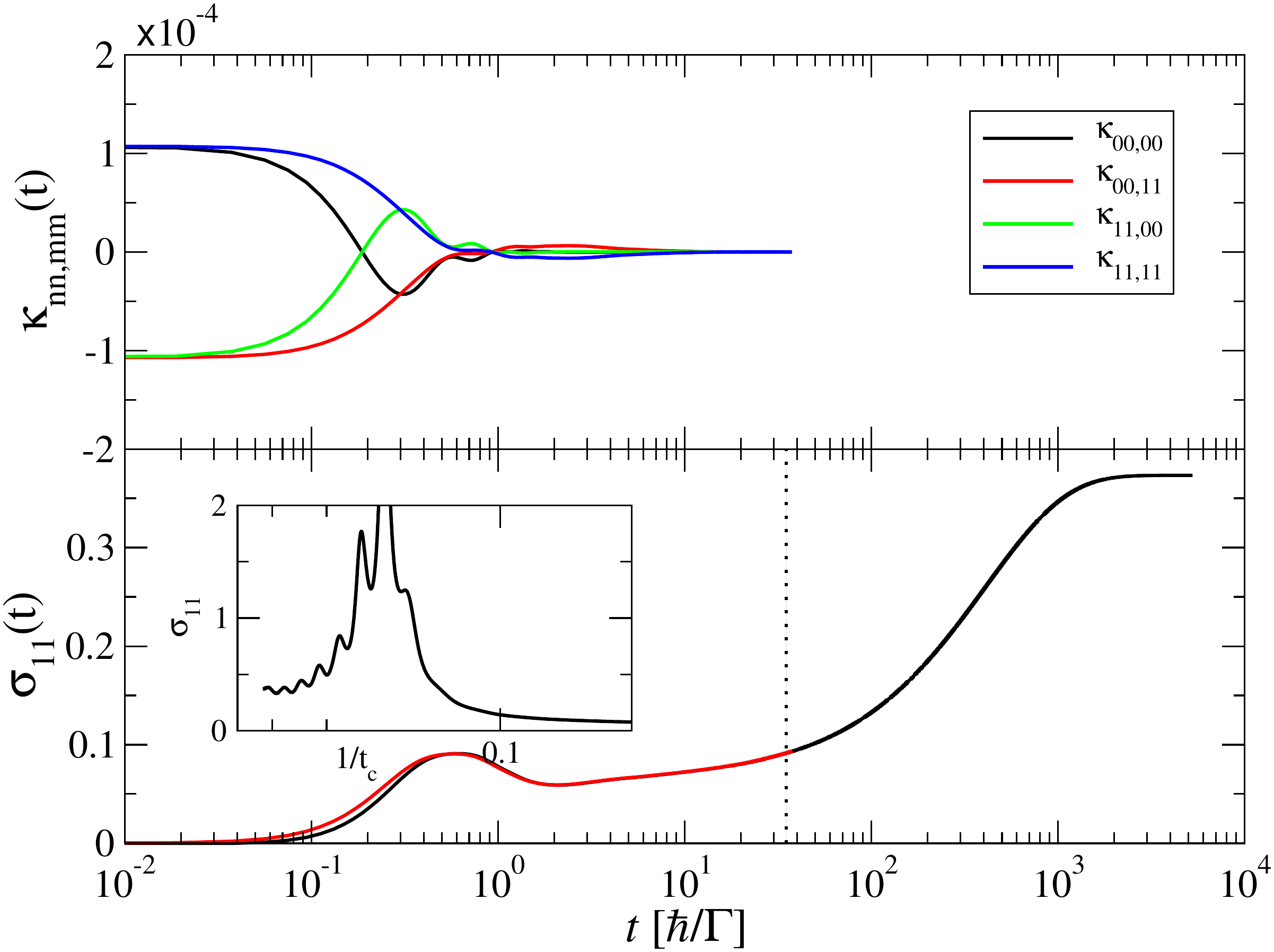} \caption{Upper panel: The elements of the memory kernel for the extended Holstein
model for $\varepsilon_{d}/\Gamma=\frac{25}{8},$ $\omega_{c}=500{\rm cm^{-1}\approx0.4\Gamma/\hbar}$,
$\lambda/\Gamma=3.5$, $\mu_{L}-\mu_{R}\approx\frac{5}{8}\Gamma$,
and $T=0{\rm K}$. Lower panel: Corresponding values of $\sigma_{11}(t)$
obtained directly from the ML-MCTDH-SQR (red curve) and from the RDM
formalism (black curve). Inset: Steady state values for $\sigma_{11}$
versus $1/t_{c}$. The dashed vertical line shows the cutoff time. }

\label{fig:sigma_tc} 
\end{figure}

As an illustration of the combined RDM and ML-MCTDH-SQR approaches,
in Fig.~\ref{fig:sigma_tc} we show the four elements of the memory
kernel (upper panel) obtained for an extended Holstein model and the
corresponding average system population ($\sigma_{11}$). The time
evolution of $\sigma(t)$ clearly agrees with the direct calculation
based on the ML-MCTDH-SQR result up to the cutoff time $t_{c}\approx35\frac{\hbar}{\Gamma}$
Beyond this time, it is difficult to converge the direct ML-MCTDH-SQR
calculations and the RDM formalism employing the memory kernel obtained
using ML-MCTDH-SQR is employed. The results obtained with the RDM
formalism shows a pronounced dynamical effect beyond $t_{c}.$ The
inset in Fig.~\ref{fig:sigma_tc} shows the steady-state value of
$\sigma_{11}$ as a function of the inverse cutoff time. As $1/t_{c}\rightarrow0$
we observe a plateau for $\sigma_{11}$ suggesting that the memory
has sufficiently decayed to $0$.

\subsubsection{Time-dependent nonequilibrium Green's function approach within the
two-time self-consistent Born approximation}

In situations where the calculation of the RDM does not converge within
the cutoff time accessible by the ML-MCTDH-SQR approach we obtain
the memory kernel from a nonequilibrium Green function approach within
the self-consistent Born approximation (SCBA). This approach is accurate
only for the perturbative regime, \textit{i.e.}, when $\lambda/\Gamma$
is small.\cite{Mitra2004} In this regime, the NEGF-SCBA expands the
cutoff time by nearly a factor of $3$, thereby providing a valuable
tool to converge the memory kernel and the RDM for weak electron-phonon
couplings.

Most applications based on NEGF within the SCBA have addressed steady-state
properties alone. Naturally, for nonequilibrium conditions one requires
a two-time representation of the Green functions (GFs), significantly
complicating the calculations. If one wishes to refrain from adopting
any type of time-local approximation,\cite{Albrecht2013} the two-time
representation limits the timescales that can be addressed directly
by the NEGF formulation. Therefore, to obtain the dynamic response
on all relevant timescales, the two-time NEGF formalism must be coupled
with the RDM formalism.

Here, we extended the two-time NEGF approach to calculate the time-dependent
left and right currents, obtain the memory kernel and the corresponding
RDM. As far as we know, the present work is also the first application
of the two-time NEGF formalism to the extended Holstein model. For
completeness, we provide a full description of the two-time NEGF approach.
We begin by introducing contour order two-time GFs~\cite{Jauho1994}
\begin{equation}
\mathcal{G}\left(t,\tau\right)=-\frac{i}{\hbar}\left\langle T_{c}d\left(t\right)d^{\dagger}\left(\tau\right)\right\rangle \label{eq:G}
\end{equation}
for the system, and 
\begin{equation}
\mathcal{D}_{\alpha}\left(t,\tau\right)=-\frac{i}{\hbar}\left\langle T_{c}x_{\alpha}\left(t\right)x_{\alpha}\left(\tau\right)\right\rangle \label{eq:D}
\end{equation}
for phonon mode $\alpha$, where $x_{\alpha}=\frac{1}{\sqrt{2}}\left(b_{\alpha}+b_{\alpha}^{\dagger}\right)$
is the phonon dimensionless coordinate, and $T_{c}$ is the Keldysh
contour time-ordering operator. We ignore correlations between different
phonon modes, \textit{i.e.} we assume $\mathcal{D}_{\alpha\beta}\left(t,\tau\right)=-\frac{i}{\hbar}\left\langle T_{c}x_{\alpha}\left(t\right)x_{\beta}\left(\tau\right)\right\rangle =0$,
if $\beta\ne\alpha$. As will become apparent below, this approximation
works quite well and is essential to describe a realistic size of
the phonon bath within the two-time formalism. The GFs in Eqs.~(\ref{eq:G})
and (\ref{eq:D}) obey the Dyson equation: 
\begin{align*}
\mathcal{G}\left(t,\tau\right) & =\mathcal{G}_{0}\left(t-\tau\right)+\\
 & \int_{c}\mbox{d}s_{1}\mbox{d}s_{2}\mathcal{G}_{0}\left(t-s_{1}\right)\Sigma\left(s_{1},s_{2}\right)\mathcal{G}\left(s_{2},\tau\right)
\end{align*}
\begin{align}
\mathcal{D}_{\alpha}\left(t,\tau\right) & =\mathcal{D}_{0\alpha}\left(t-\tau\right)+\nonumber \\
 & \int_{c}\mbox{d}s_{1}\mbox{d}s_{2}\mathcal{D}_{0\alpha}\left(t-s_{1}\right)\Pi_{\alpha}\left(s_{1},s_{2}\right)\mathcal{D_{\alpha}}\left(s_{2},\tau\right)\label{eq:dyson}
\end{align}
where $\mathcal{G}_{0}\left(t\right)$ and $\mathcal{D}_{0\alpha}\left(t\right)$
are the bare propagators of the electronic degrees of freedom on the
quantum dot and phonon mode $\alpha$, respectively, evolving under
$H_{s}+H_{{\rm ph}}$, and $\int_{c}$ is a time integration on the
Keldysh contour. In the above, $\Sigma$ and $\Pi_{\alpha}$ are the
system and phonon self-energies, respectively. As pointed out above,
we apply the SCBA to obtain these self-energies, which correspond
to a partial summation of the diagrams beyond the simpler second order
approximation where each bare GF is replaced by the full propagator.
A self-consistence solution is computational far more demanding, but
leads to a results which is more satisfactory from a theoretical point
of view. In fact, we find that the SCBA is accurate even for electron-phonon
couplings of the order of $\lambda/\Gamma\approx3$, slightly outside
the perturbative regime. Within the SCBA, the system and phonon self-energies
are given by 
\begin{equation}
\Sigma\left(t,\tau\right)=\Sigma_{\ell}\left(t-\tau\right)+i\hbar\sum_{\alpha}M_{\alpha}^{2}\mathcal{D}_{\alpha}\left(t,\tau\right)\mathcal{G}\left(t,\tau\right)\label{eq:sigma_r}
\end{equation}
and 
\begin{eqnarray}
\Pi_{\alpha}\left(t,\tau\right) & = & -i\hbar M_{\alpha}^{2}\mathcal{G}\left(t,\tau\right)\mathcal{G}\left(\tau,t\right),\label{eq:Pi_r}
\end{eqnarray}
respectively. In the above expression, we neglected virtual processes
coupling different phonon modes contributing to the self-energies.
$\Sigma_{\ell}\left(t\right)=\Sigma_{\ell,L}\left(t\right)+\Sigma_{\ell,R}\left(t\right)$
represents the self-energy arising from the coupling to the leads,
with retarded ('$r$') and lesser ('$<$') self-energies defined by
$i\Sigma_{\ell,L/R}^{r}(t)=\frac{1}{2\pi}\int\Gamma_{L/R}\left(\varepsilon\right)e^{-\frac{i}{\hbar}\varepsilon t}\mbox{d}\varepsilon$
and $i\Sigma_{\ell,L/R}^{<}(t)=-\frac{1}{2\pi}\int\Gamma_{L/R}\left(\varepsilon\right)f\left(\varepsilon-\mu_{L/R}\right)e^{-\frac{i}{\hbar}\varepsilon t}\mbox{d}\varepsilon$
, respectively, and $f\left(\varepsilon\right)$ is the Fermi-Dirac
distribution. These Keldysh GFs and self-energies are obtained using
Langreth rules.\cite{Haug2008,stefanucci_nonequilibrium_2013}

Once the expressions for the self-energies are given, we seek a solution
for the two-time GFs. Instead of solving the usual Dyson equations,
a simple Leibniz rule can be applied to reduce these equations to
the Kadanoff-Baym form.\cite{kadanoff_quantum_1994,myohanen_kadanoff-baym_2010,stefanucci_nonequilibrium_2013}
For the retarded GFs, this reads 
\begin{align}
i\hbar\frac{\partial\mathcal{G}^{r}\left(t,\tau\right)}{\partial t} & =\delta\left(t-\tau\right)+\varepsilon_{d}\mathcal{G}^{r}\left(t,\tau\right)\nonumber \\
 & +\int_{\tau}^{t}\Sigma^{R}\left(t,s\right)\mathcal{G}^{r}\left(s,\tau\right)\mbox{d}s,\label{eq:gr(t)}
\end{align}

\begin{eqnarray}
\frac{\partial^{2}\mathcal{D}_{\alpha}^{r}\left(t,\tau\right)}{\partial t^{2}} & = & -\frac{2\omega_{\alpha}}{\hbar}\delta\left(t-\tau\right)-\omega_{\alpha}^{2}\mathcal{D}_{\alpha}^{r}\left(t,\tau\right)\nonumber \\
 & - & \frac{2\omega_{\alpha}}{\hbar}\int_{\tau}^{t}\Pi_{\alpha}^{r}\left(t,s\right)\mathcal{D}_{\alpha}^{r}\left(s,\tau\right)\mbox{d}s\label{eq:D_r}
\end{eqnarray}
and for the lesser GFs one finds: 
\begin{eqnarray}
i\hbar\frac{\partial\mathcal{G}^{<}\left(t,\tau\right)}{\partial t} & = & \varepsilon_{d}\mathcal{G}^{<}\left(t,\tau\right)+\int_{0}^{t}\Sigma^{r}\left(t,s\right)\mathcal{G}^{<}\left(s,\tau\right)\mbox{d}s\nonumber \\
 & + & \int_{0}^{\tau}\Sigma^{<}\left(t,s\right)\left(\mathcal{G}^{r}\left(\tau,s\right)\right)^{\dagger}\mbox{d}s,\label{eq:KB_G_L}
\end{eqnarray}
\begin{eqnarray}
\frac{\partial^{2}\mathcal{D}_{\alpha}^{<}\left(t,\tau\right)}{\partial t^{2}} & = & -\omega_{\alpha}^{2}\mathcal{D}_{\alpha}^{<}\left(t,\tau\right)\nonumber \\
 &  & -\frac{2\omega_{\alpha}}{\hbar}\int_{0}^{t}\Pi_{\alpha}^{r}\left(t,s\right)\mathcal{D}_{\alpha}^{<}\left(s,\tau\right)\mbox{d}s\nonumber \\
 & - & \frac{2\omega_{\alpha}}{\hbar}\int_{0}^{\tau}\Pi_{\alpha}^{<}\left(t,s\right)\left(\mathcal{D}_{\alpha}^{r}\left(\tau,s\right)\right)^{\dagger}\mbox{d}s.\label{eq:D_l}
\end{eqnarray}
The left and right currents can be obtained from the Meir-Wingreen
formula~\cite{meir_landauer_1992}: 
\begin{eqnarray}
I_{m}^{L,R}(t) & = & -\frac{2e}{\hbar}\Im\left\{ \int_{0}^{t}\mathcal{G}^{<}\left(t,s\right)i\Sigma_{\ell,L/R}^{r}(t-s)\mbox{d}s\right.\nonumber \\
 & + & \left.\int_{0}^{t}\mathcal{G}^{r}\left(t,s\right)i\Sigma_{\ell,L/R}^{<}(t-s)\mbox{d}s\right\} .\nonumber \\
\label{eq:left_current}
\end{eqnarray}
Here $m$ denotes the dependence on the initial preparation, which
enters through the initial values taken for $\mathcal{G}_{0}^{<}\left(0\right)=-\frac{i}{\hbar}\left\langle m\right|d^{\dagger}\left(0\right)d\left(0\right)\left|m\right\rangle =-\frac{i}{\hbar}\cdot m$.

\subsection{Initial conditions}

\begin{figure*}[t]
\includegraphics[width=8cm]{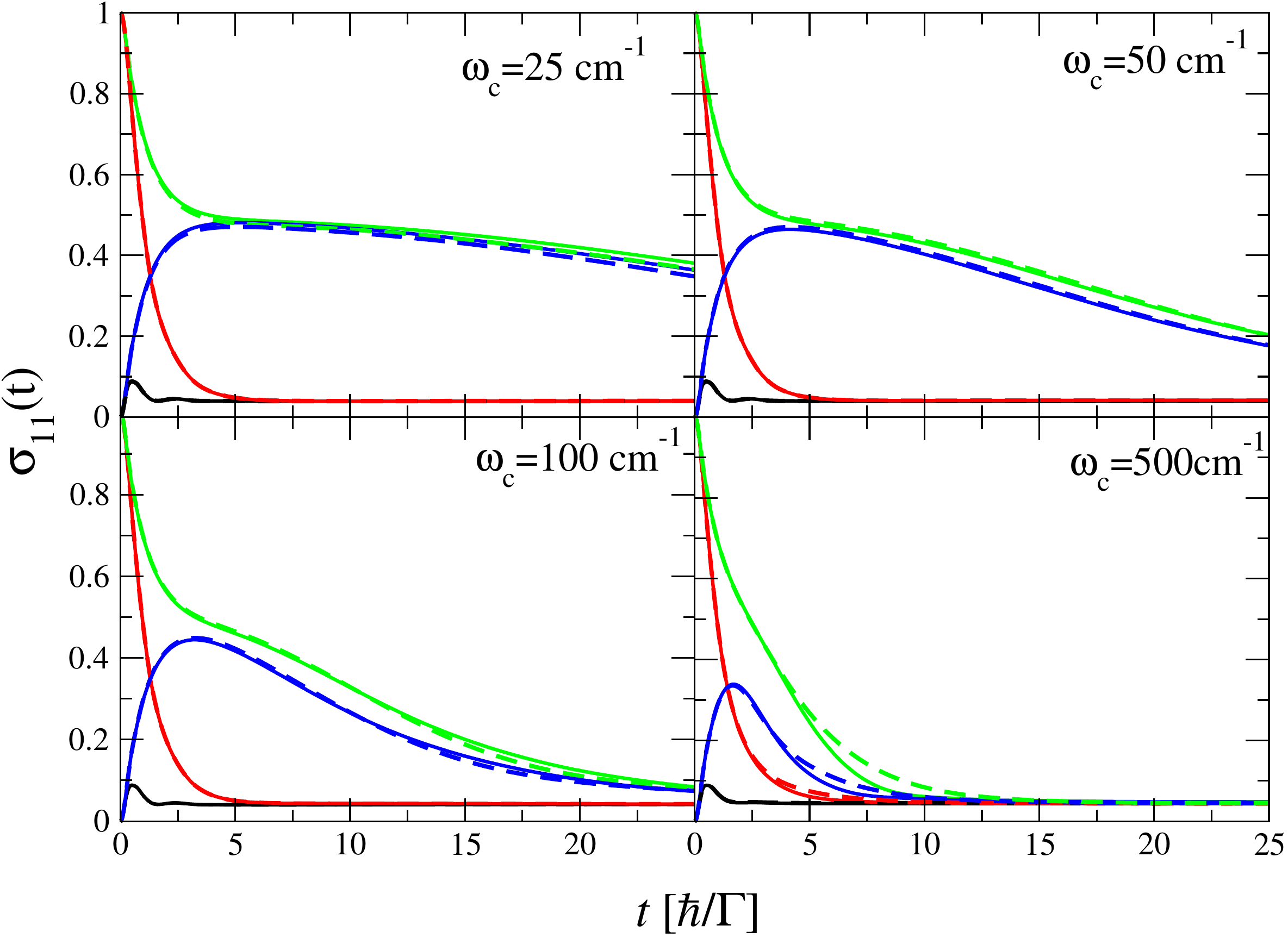}\includegraphics[width=8cm]{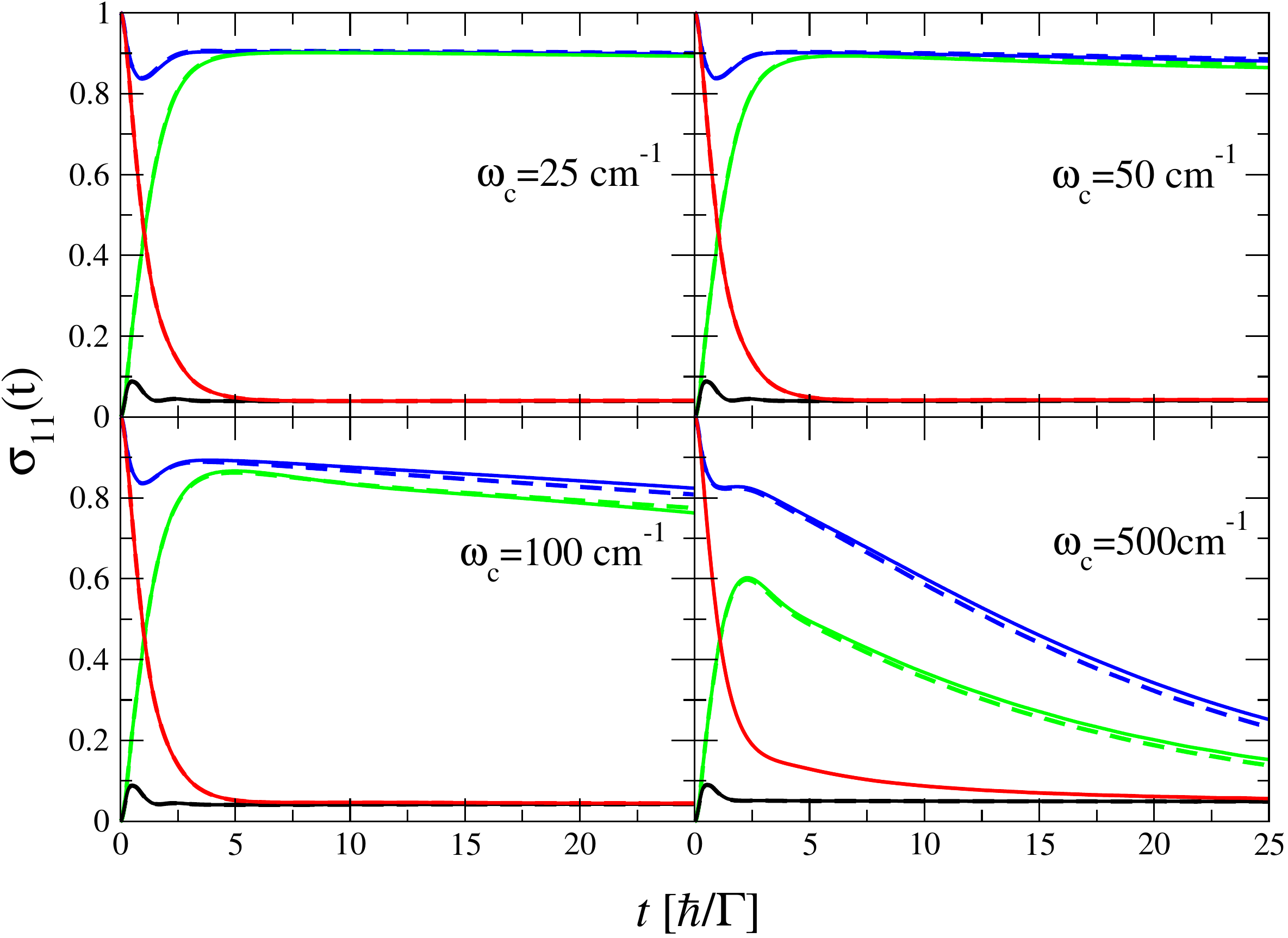}
\caption{Comparison of the dot population ($\sigma_{11}(t)$) obtained from
the ML-MCTDH-SQR (solid lines) and NEGF-SCBA (dashed lines) approaches
for $\varepsilon_{d}/\Gamma=\frac{25}{8}$, $\lambda/\Gamma\approx1.5$
(left panels) and $\lambda/\Gamma\approx2.3$ (right panels), for
frequencies in the range of $25-500\mbox{cm\ensuremath{^{-1}}}$ ($\approx0.02-0.4$
in units of $\Gamma/\hbar$). The different curves correspond to different
initial conditions: Black - unoccupied with $\delta_{\alpha}=0$,
\textcolor{red}{Red} - occupied with $\delta_{\alpha}=0$, \textcolor{blue}{Blue}
- unoccupied with $\delta_{\alpha}=1$ and \textcolor{green}{Green}
- occupied with $\delta_{\alpha}=1$. }

\label{fig:comparison} 
\end{figure*}

To characterize the population dynamics, we must define the initial
condition for the full density matrix of the system and bath. To simplify
the description within the RDM formalism, we start with a factorized
initial condition, which implies that $\vartheta(t)$ in Eq.~(\ref{eq:theta(t)})
vanishes for all times. The initial density matrix, $\rho(0)$, is
given by 
\begin{equation}
\rho\left(0\right)=\sigma\left(0\right)\otimes\rho_{B}\left(0\right)=\sigma\left(0\right)\otimes\rho_{\mbox{ph}}\left(0\right)\otimes\rho_{\ell}^{L}(0)\otimes\rho_{\ell}^{R}(0),\label{eq:rho0}
\end{equation}
where $\sigma\left(0\right)$ determines whether the electronic level
is initially occupied/unoccupied, 
\begin{equation}
\rho_{\ell}^{L/R}(0)=\exp\left[-\beta\left(\sum_{k\in L/R}\left(\varepsilon_{k}-\mu_{L/R}\right)a_{k}^{\dagger}a_{k}\right)\right],\label{eq:rholeads}
\end{equation}
is the initial density matrix for the leads, and

\begin{align}
\rho_{\mbox{ph}}\left(0\right) & =\exp\left[-\beta\left\{ \sum_{\alpha}\hbar\omega_{\alpha}\left(b_{\alpha}^{\dagger}b_{\alpha}+\frac{1}{2}\right)\right.\right.\nonumber \\
 & \left.\left.+\sum_{\alpha}\delta_{\alpha}M_{\alpha}\left(b_{\alpha}^{\dagger}+b_{\alpha}\right)\right\} \right]\label{eq:rhoph}
\end{align}
represents the initial density matrix of the phonon bath. In the above
equations $\beta=\frac{1}{k_{B}T}$ is the inverse temperature.

The calculation of the different elements of the memory kernel require
the calculation of the current for different initial occupation of
the system ($I_{m}^{L,R}(t)$), \textit{i.e}., for different values
of $\sigma\left(0\right)$. For the ML-MCTDH-SQR approach this amounts
to selecting different initial single-particle wave functions for
the system while, as pointed above, for the NEGF, the only term that
depends on the initial electronic preparation of the system is $\mathcal{G}_{0}^{<}\left(0\right)$.
It has been shown that for the extended Holstein model, the steady
state values of $\sigma$ are independent of the choice $\sigma(0)$~\cite{wilner_bistability_2013},
\textit{i.e.}, the choice of $\mathcal{G}_{0}^{<}\left(0\right)$,
but the dynamic response and relaxation to steady-state does depend
on $\sigma(0)$.

We will also consider two different initial conditions for the phonons,
one where $\delta_{\alpha}=0$ in Eq.~(\ref{eq:rhoph}) corresponding
to phonons initially equilibrated with an unoccupied dot, and another
where $\delta_{\alpha}=1$ corresponding to phonons equilibrated to
an occupied dot. Again, the description of these two initial conditions
is rather simple within the ML-MCTDH-SQR approach and one selects
the initial phonon wave function to correspond to one of these initial
conditions. Within the NEGF formalism, this is a bit more delicate.
The phonon initial condition enters the Kadanoff-Baym equations through
the equitime lesser bare phonon GF, ${\cal D}_{0,\alpha}^{<}(0)$.
For $\delta_{\alpha}=0$ we set ${\cal D}_{0,\alpha}^{<}(0)=-\frac{i}{\hbar}\left(2n\left(\hbar\omega_{\alpha}\right)+1\right)$,
where $n(\omega)=\frac{1}{e^{\beta\omega}-1}$ is the Bose-Einstein
distribution.

For $\delta_{\alpha}=1$ one can use a similar strategy and determine
${\cal D}_{0,\alpha}^{<}(0)$ according to Eq.~(\ref{eq:rhoph}).
However, this would lead to large deviations of the NEGF approach
from the numerically exact ML-MCTDH-SQR results, since this initial
condition amounts to a situation where the phonons are equilibrated
in the well corresponding to the occupied dot, a situation far from
the perturbative regime about which the NEGF equations where derived.
To resolve this and provide an equally accurate description of the
NEGF-SCBA for the shifted phonon distribution, we propose to transform
the phonon Hamiltonian in Eq.(\ref{eq:Hamiltonian-phonons}) by redefining
a set of shifted ladder operators $\tilde{b}_{\alpha}=b_{\alpha}+\frac{M_{\alpha}}{\hbar\omega_{\alpha}}$
combined with particle/hole transformation $d\rightarrow\tilde{d}^{\dagger},d^{\dagger}\rightarrow\tilde{d}$.
With that, the shifted phonon Hamiltonian is given by: 
\begin{eqnarray}
H^{\delta=1} & = & \left(2\lambda-\varepsilon_{d}\right)\tilde{d}^{\dagger}\tilde{d}+\sum_{k\in L,R}\varepsilon_{k}a_{k}^{\dagger}a_{k}\nonumber \\
 & + & \underset{\alpha}{\sum}\hbar\omega_{\alpha}\left(\tilde{b}_{\alpha}^{\dagger}\tilde{b}_{\alpha}+\frac{1}{2}\right)+\sum_{k\in L,R}\left(t_{k}\tilde{d}^{\dagger}a_{k}^{\dagger}+t_{k}^{*}\tilde{d}a_{k}\right)\nonumber \\
 & - & \tilde{d}^{\dagger}\tilde{d}\sum_{\alpha}M_{\alpha}\left(\tilde{b}_{\alpha}^{\dagger}+\tilde{b}_{\alpha}\right),\label{eq:shifted_Hamiltonian}
\end{eqnarray}
which is identical to the phonon Hamiltonian in Eq.(\ref{eq:Hamiltonian-phonons})
with $\varepsilon_{d}\rightarrow2\lambda-\varepsilon_{d}$ and $M_{\alpha}\rightarrow-M_{\alpha}$.
Thus, one can adopt the NEGF-SCBA equations derived above with parameters
reflecting this transformation. The initial condition for the shifted
phonons will now correspond to $\left\langle \tilde{d}^{\dagger}\tilde{d}\right\rangle =0$.
In practice we use the NEGF-SCBA equations for both initial conditions
of the phonons with the original set of parameters and ${\cal D}_{0,\alpha}^{<}(0)=-\frac{i}{\hbar}\left(2n\left(\hbar\omega_{\alpha}\right)+1\right)$
for $\delta_{\alpha}=0$ and with $\varepsilon_{d}\rightarrow2\lambda-\varepsilon_{d}$
, $M_{\alpha}\rightarrow-M_{\alpha}$ for $\delta_{\alpha}=1$ with
the same values for ${\cal D}_{0,\alpha}^{<}(0)$.

In Fig.~\ref{fig:comparison} we compare the short-time behavior
of the RDM obtained from the NEGF-SCBA to the numerically converged
ML-MCTDH-SQR approach. Four initial preparations of the system were
considered at different values of $\lambda$ and $\omega_{c}.$ The
agreement between the NEGF-SCBA and the ML-MCTDH-SQR results is remarkable
even slightly outside the perturbative regime by which the SCBA is
expected to fail, \textit{i.e.}, for $\lambda/\Gamma>1$.\cite{Mitra2004}
While the ML-MCTDH-SQR is limited to times of the order of $35\frac{\hbar}{\Gamma}$
the NEGF-SCBA can be used (within our computational resources) to
times of the order of $100\frac{\hbar}{\Gamma}$, which as shown below,
is necessary to converge the RDM to steady-state for certain parameters.
We note in passing that for values of $\lambda/\Gamma>3$ we find
that the NEGF-SCBA shows a pronounced deviation from the numerically
converged results and thus can only provide a qualitative picture.
However, for $\lambda/\Gamma<\frac{5}{2}$ it seems safe to use the
NEGF-SCBA approach.

\begin{figure*}[t]
\includegraphics[width=16cm]{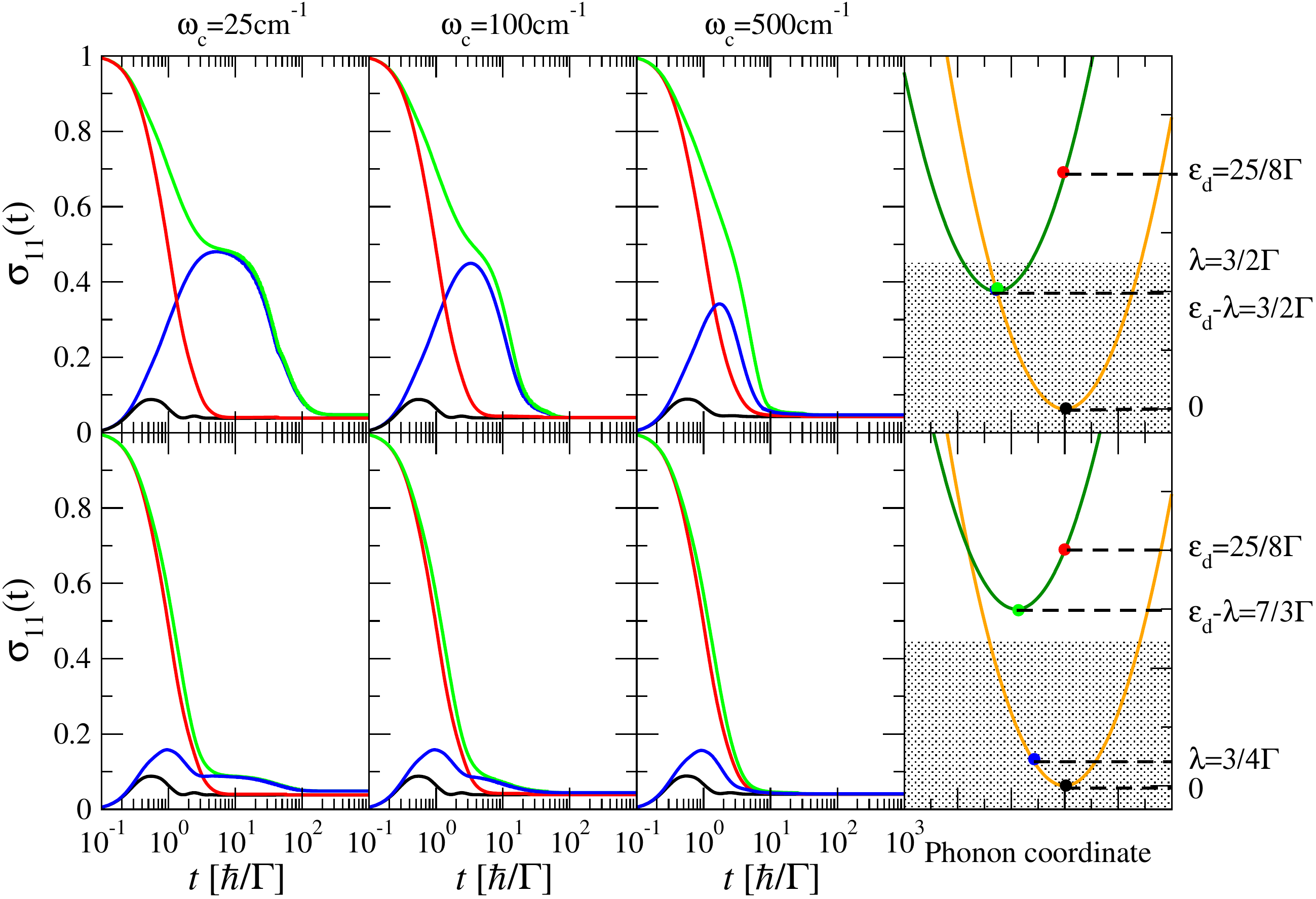} \caption{Left matrix-panels: Dot population ($\sigma_{11}(t)$) obtained from
the RDM combined with the NEGF-SCBA for $\lambda/\Gamma=\frac{3}{2}$
(upper row panels) and $\lambda/\Gamma=\frac{3}{4}$ (lower row panels)
for frequencies in the range of $25-500\mbox{cm\ensuremath{^{-1}}}$
($\approx0.02-0.4$ in units of $\Gamma/\hbar$). Black, red, blue,
and green curves correspond to unoccupied / occupied and $\delta_{\alpha}=0$
/ $\delta_{\alpha}=1$, respectively. Right column-panels: Schematic
sketch of the diabatic harmonic potential energy surfaces of the neutral
(orange) and charged (green) state for the two values of $\lambda$.
The marked values to the right label the dot and phonon minimum energy
corresponding to each of the four initial condition. For each initial
condition, we label the minimum energy with a solid circle with matching
colors. Shaded area corresponds to the applied source-drain bias.}

\label{fig:short-time} 
\end{figure*}

\section{Analysis of the nonequilibrium dynamics at different time scales\label{sec:Time-Scales-Analysis}}

The nonequilibrium dynamics of the quantum dot, represented by the
RDM, exhibits various timescales, which are analyzed in this section
using the approaches introduced above. We first consider the dynamics
for relatively short times, \textit{i.e.}, on timescales characterized
by the dot-lead coupling ($\tau_{\ell}\approx\frac{\hbar}{\Gamma}$)
and the typical phonon frequency $\tau_{{\rm ph}}\approx\frac{1}{\omega_{c}}$.
We show that the appearance of rapid decays of the RDM to steady state
depends also the specific model parameters, in particular whether
the coupling to the phonons shifts the energy of the dot in or out
of the bias window, which is defined by the chemical potentials of
the two leads. Next, we study the long-time decay of the RDM to steady
state and address both the adiabatic\textbf{ }(\textbf{$\hbar\omega_{c}\ll\Gamma$)}
and non-adiabatic limits. In all results presented below, we consider
the low temperature limit ($T=0$).

\subsection{Short and intermediate time scales}

In Fig.~\ref{fig:short-time}, we plot the average dot population
given by the diagonal occupied element of the RDM ($\sigma_{11}(t)$)
for several typical phonon frequencies, for two values of the reorganization
energy, $\lambda/\Gamma\approx\frac{3}{4}$ (lower panel) and $\lambda/\Gamma\approx\frac{3}{2}$
(upper panel), for $\mu_{L}=-\mu_{R}\approx\frac{1}{3}\Gamma$. As
shown above, this regime of electron-phonon coupling is well suited
for the two-time NEGF-SCBA combined with the RDM formalism. We consider
four different initial preparations of the dot and phonon density
matrices: Occupied/empty dot where $\sigma(0)=\left(\begin{array}{cc}
0 & 0\\
0 & 1
\end{array}\right)$ for an occupied dot and $\sigma(0)=\left(\begin{array}{cc}
1 & 0\\
0 & 0
\end{array}\right)$ otherwise, and shifted/unshifted phonons with $\delta_{\alpha}=1,0$,
respectively. In all cases shown, the dot population decays to the
same steady-state value, regardless of the initial preparation of
the dot/phonons. For the case of $\delta_{\alpha}=0$ (black and red
curves) we find that the dynamics are characterized by a single timescale
governed by $\tau_{\ell}\approx\frac{\hbar}{\Gamma}$. For $\delta_{\alpha}=1$
(blue and green curves), this initial transient is followed by a decay
on timescales of $\tau_{{\rm ph}}\approx\frac{1}{\omega_{c}}$ for
the larger reorganization energy (upper panels) while for $\lambda/\Gamma\approx\frac{3}{4}$
the phonon frequency is not always noticeable (lower panels). For
higher values of $\lambda$ not shown in Fig.~\ref{fig:short-time},
the picture will reverse, namely, dynamics on timescales of $\tau_{{\rm ph}}$
will appear for an initially unshifted phonon distribution.

\begin{figure*}[t]
\includegraphics[width=16cm]{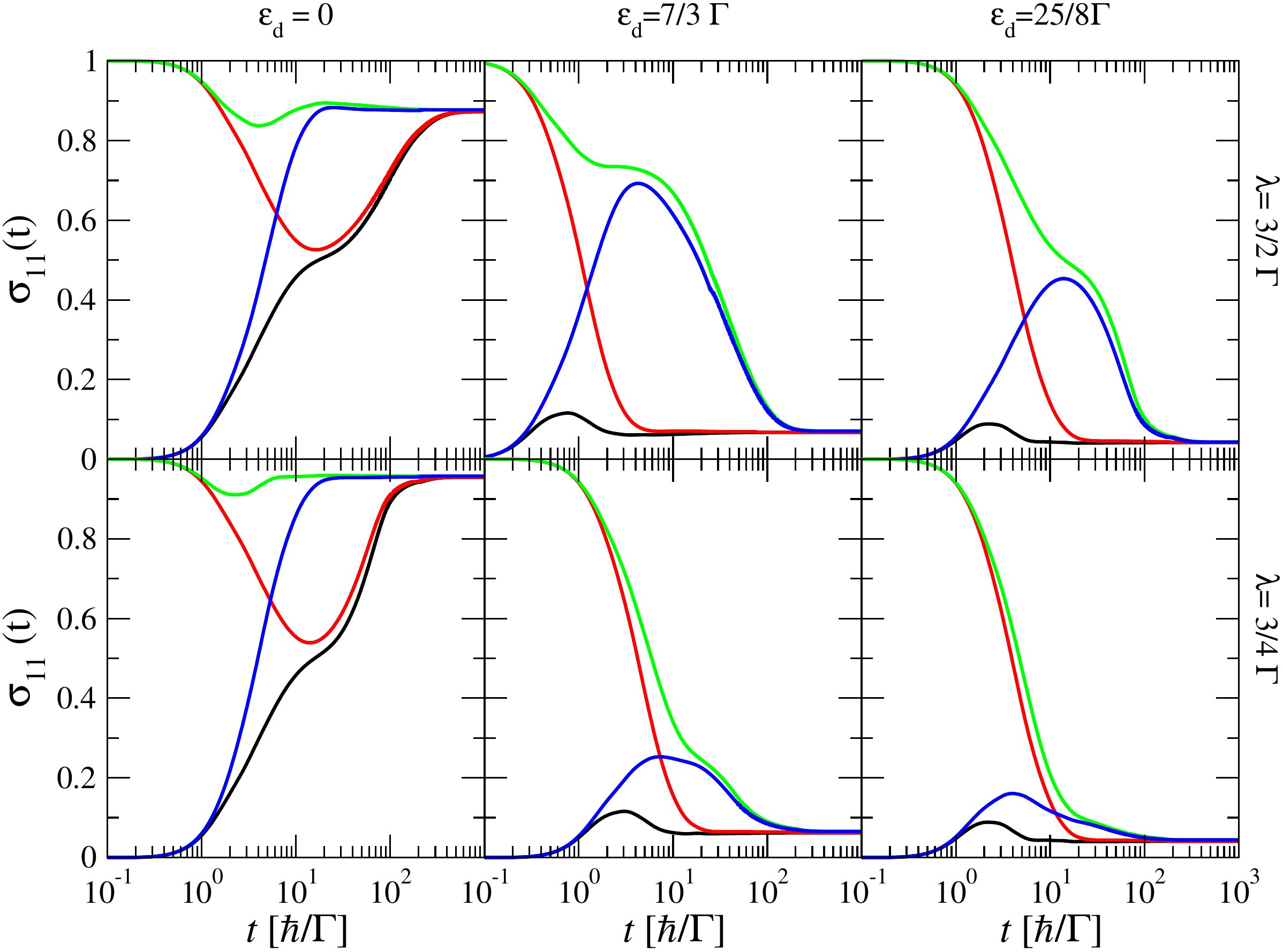} \caption{Dot population ($\sigma_{11}(t)$) obtained from the RDM combined
with the NEGF-SCBA for $\omega_{c}=100\mbox{cm\ensuremath{^{-1}\approx0.08\Gamma/\hbar}}$.
Black, red, blue, and green curves correspond to unoccupied / occupied
and $\delta_{\alpha}=0$ / $\delta_{\alpha}=1$, respectively.}

\label{fig:short-time-bias} 
\end{figure*}

To better understand the intermediate time behavior, we provide a
sketch of the two diabatic potential energy surfaces for a typical
phonon frequency of $\omega_{c}=100\mbox{cm}^{-1}$ for the two values
of $\lambda$. For each plot, we also indicate the sum of dot and
phonon energy of the $4$ different initial conditions. It is quite
clear that the most stable configuration is that of an empty dot with
an unshifted phonon ($\delta_{\alpha}=0$), which for small bias-voltages,
would likely be the steady-state configuration. Therefore, regardless
of the value of $\lambda$, when the system initial phonon distribution
corresponds to the unshifted case (black and red curves), the phonons
are already close to their steady-state distribution and the dynamics
of the RDM are governed by the electronic decay determined by the
coupling to the leads ($\Gamma$).

Considering the case of $\lambda/\Gamma=\frac{3}{2}$ for the shifted
initial phonon distribution, at short times ($\tau_{\ell}$) the population
of the dot decreases or increases to a value of $\frac{1}{2}$, depending
on whether the dot was occupied or empty initially, respectively.
To understand this, we define the instantaneous difference in energy
between an occupied and empty dot as $\delta\varepsilon$. For $x=0$
(the minimum of the unshifted well) $\delta\varepsilon=\varepsilon_{d}$
and for $x=-\sqrt{2}\frac{M}{\hbar\omega_{c}}$ (the minimum of the
shifted well) $\delta\varepsilon=\varepsilon_{d}-2\lambda$. Returning
to the case $\lambda/\Gamma=\frac{3}{2}$ for the shifted initial
phonon distribution, $\delta\varepsilon=\varepsilon_{d}-2\lambda\approx0$
is nearly at the symmetric point about the bias window of conduction.
Thus, freezing the phonons would lead to a steady-state population
close to $\frac{1}{2},$which is indeed observed for times $\tau_{\ell}<t<\tau_{{\rm ph}}$
where the dot population levels at $\approx\frac{1}{2}$. The phonons,
of course, are not frozen and as the system relaxes to the more stable
well on timescales given by $\tau_{{\rm ph}}$. During this process,
the instantaneous value of $\delta\varepsilon$ shifts above the bias
conduction windows, resulting in a decay of the dot population.

\begin{figure*}[t]
\includegraphics[width=16cm]{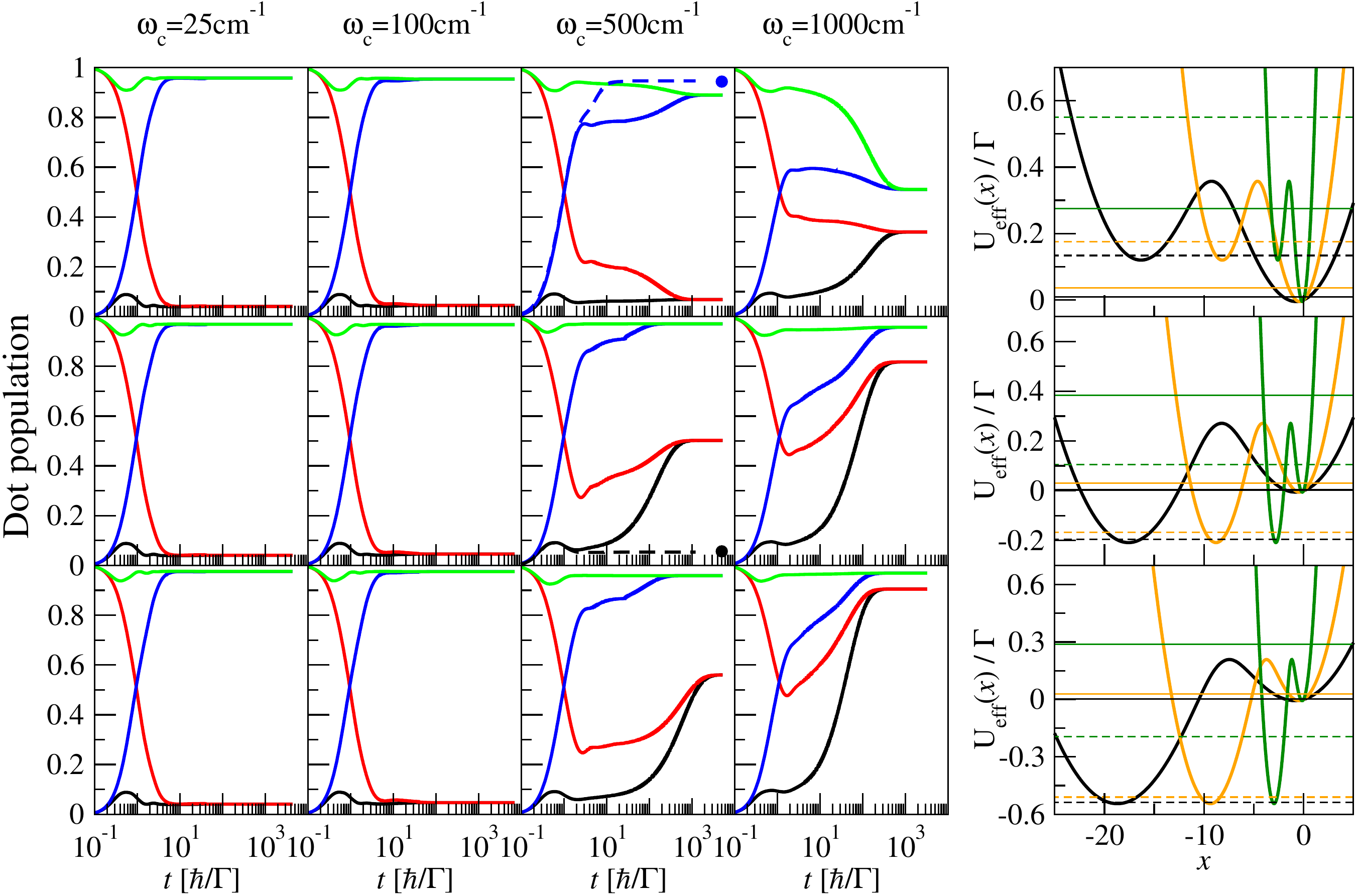} \caption{Left matrix-panels: Dot population obtained from the RDM formalism
combined with the ML-MCTDH-SQR approach for $\lambda/\Gamma=3.1$
(upper row panels), $\lambda/\Gamma=3.5$ (middle row panels), and
$\lambda/\Gamma=3.9$ (lower row panels) for frequencies in the range
of $25-1000\mbox{cm\ensuremath{^{-1}}}$ ($\approx0.02-0.8$ in units
of $\Gamma/\hbar$). Black, red, blue, and green curves correspond
to unoccupied / occupied and $\delta_{\alpha}=0$ / $\delta_{\alpha}=1$,
respectively. Dashed curves were obtained from the RDM formalism combined
with the NEGF-SCBA and the solid circles are the NEGF-SCBA steady-state
values. Right column panels: The effective potentials corresponding
to each value of $\lambda$. Black, orange and deep green curves are
for $\omega_{c}=25$, $100$, and $1000\mbox{cm\ensuremath{^{-1}}}$,
respectively. The horizontal solid and dashed lines represent the
ground state energy of the right and left well, respectively.}

\label{fig:long-time} 
\end{figure*}

For the smaller reorganization energy ($\lambda/\Gamma=\frac{3}{4}$)
the energy difference $\delta\varepsilon$ is well above the bias
window of conduction, and thus, the population of the dot never levels
at values typical of resonance situations. Inevitably, the system
will relax to the more stable well corresponding to $\delta_{\alpha}=0$
on a timescale $\tau_{{\rm ph}}$. Whether this appears in the dynamics
of the RDM depends on the value of the dot population. For non-vanishing
$\sigma_{11}(t)$, a clear signature of $\tau_{{\rm ph}}$ is still
evident.

The picture that emerges is rather simple. At short times, the dynamics
of the RDM is always characterized by the coupling to the leads as
long as $\hbar\omega_{c}<\Gamma$. The appearance of an additional
timescale ($\tau_{{\rm ph}}$) depends on whether the phonons are
initial equilibrated at the more stable well or not, and also whether
the instantaneous energy difference between the occupied and empty
dot passes through the bias conduction window as the system relaxes
to steady-state. To further support this we show in Fig.~\ref{fig:short-time-bias}
results for the dot population for different values of $\varepsilon_{d}$
and a higher bias voltage $\mu_{L}=-\mu_{R}\approx\frac{2}{3}\Gamma$,
for the same values of $\lambda$. The two left panels show results
for $\varepsilon_{d}=0$ in which the shifted well is the more stable
one. As clearly evident, the role of the different initial conditions
is reversed and the dynamics of the RDM corresponding to the shifted
initial condition relax rapidly to the steady-state while the case
of the unshifted initial condition show intermediate transient behavior
(with dot population approaching $\frac{1}{2}$ since $\delta\varepsilon=0$
for this case) with a characteristic timescale $\tau_{{\rm ph}}$.

The case of $\varepsilon_{d}=\frac{7}{3}\Gamma$ and $\lambda=\frac{3}{2}\Gamma$
is special since $\delta\varepsilon=\varepsilon_{d}-2\lambda=-\frac{2}{3}\Gamma$
equals to the lower conduction edge ($\mu_{R}=-\frac{2}{3}\Gamma$).
As the system relaxes to the stable well the instantaneous value of
$\delta\varepsilon$ scans the entire bias conduction window and the
population of the dot increases above $\frac{1}{2}$, as it should
for asymmetric resonant situations. When $\delta\varepsilon$ increases
above the upper conduction edge, the dot population decreases with
a typical timescale equal to $\tau_{{\rm ph}}$. This increase of
the dot population above $\frac{1}{2}$ is not observed for $\lambda=\frac{3}{4}\Gamma$,
since for this case $\delta\varepsilon=\frac{5}{6}$ is slightly above
the upper conduction edge, and the system is never at resonance throughout
the dynamics. This explains the lower values of the dot population
at intermediate times.

\subsection{Dynamics on longer time scales induced by electron-phonon interaction}

Next, we consider the dynamics on longer time scales, induced by the
coupling between the electron and phonon degrees of freedom. In Fig.~\ref{fig:long-time}
we plot the dot population for a range of values of $\omega_{c}$
and $\lambda$, and for the four different initial conditions discussed
above. The results span the crossover between the adiabatic ($\hbar\omega_{c}\ll\Gamma$)
to the non-adiabatic ($\hbar\omega_{c}\rightarrow\Gamma$) limits.
The values of the reorganization energy chosen are somewhat above
the perturbative regime ($\lambda/\Gamma>3$) in which the NEGF-SCBA
is accurate. Therefore, we obtain the input required to generate the
memory kernel and the RDM from the ML-MCTDH-SQR approach. In all cases
shown, we used a cutoff time $t_{c}\approx25\frac{\hbar}{\Gamma}$,
sufficient to converge the rate of decay of the RDM at long times.
The value of the steady state obtained from the cutoff approximation
for these results, however, is not converged within the maximal cutoff
time used of $t_{c}\approx35\frac{\hbar}{\Gamma}$, which implies
that there maybe a longer timescale by which the system relaxes.

The two left column-panels of Fig.~\ref{fig:long-time} show results
for slow phonons ($\omega_{c}\le100\mbox{cm\ensuremath{^{-1}\approx0.08\Gamma/\hbar}}$),
i.e.\ in the adiabatic regime. For the specific choice of parameters,
we find that the long-time limit of the dot population depends on
the initial phonon distribution but not on the initial dot occupation.
The difference between the long-time plateau solutions diminishes
as the phonon frequency increases, and will eventually vanish at the
crossover to the non-adiabatic limit. The dependence of the dot occupation
for long times on the initial state suggests the existence of bistability.
This bistability has been the subject of our recent study~\cite{wilner_bistability_2013}
and previous work~\cite{galperin_hysteresis_2005,Mozyrsky05,albrecht_bistability_2012}
It will be addressed briefly later in this section and in more detail
in section~\ref{sec:bistability}.

Concerning the dynamics, we find that in the adiabatic limit the RDM
decays rapidly to a plateau, with a value that depends on the initial
phonon distribution. The RDM decay is characterized by a single timescale,
$\tau_{\ell}$, determined by the coupling to the leads. The existence
of the plateau and the plateau value are insensitive to further increasing
the cutoff time up to the limit of the ML-MCTDH-SQR approach, which
is $t_{c}\approx35\frac{\hbar}{\Gamma}$. A significantly different
behavior is observed for $\omega_{c}\ge500\mbox{cm\ensuremath{^{-1}\approx0.4\Gamma/\hbar}}$,
which is near the crossover to the non-adiabatic limit. While the
short time dynamics are very similar and are governed by the coupling
to the leads with a time scale $\tau_{\ell}$, a pronounced long time
decay is observed and then the system levels at a plateau. We note
in passing that a similar long time decay has been reported by Albrecht
\textit{et al}.~\cite{Albrecht2013} for a single-phonon Holstein
model (rather than a bath of phonons), using an NEGF approach within
a quasi-adiabatic, single-time approximation. The results shown in
Fig.~\ref{fig:long-time} are based on a numerically exact formalism,
and are therefore, free of any approximation or bias.

To understand the long time behavior in the adiabatic limit, we have
calculated the adiabatic tunneling times as well as the transition
probabilities for an effective adiabatic potential sketched in the
right column-panels of Fig.~\ref{fig:long-time}. The effective potential,

\begin{eqnarray}
U_{\mbox{eff}}\left(x\right)=U\left(x\right)+\int_{-\infty}^{x}\mbox{d}y\, n\left(y\right)\frac{\mbox{d}\varepsilon\left(y\right)}{\mbox{d}y}\label{eq:ueff}
\end{eqnarray}
is given as a sum of the bare potential $U\left(x\right)=\frac{\hbar}{2}\omega_{c}x^{2}$
and the potential of mean force, $\int_{-\infty}^{x}\mbox{d}y\, n\left(y\right)\frac{\mbox{d}\varepsilon\left(y\right)}{\mbox{d}y}$.
Here, $\varepsilon\left(x\right)=\varepsilon_{d}+\sqrt{2}M_{c}x$
is the unweighted instantaneous dot energy and

\begin{eqnarray}
n\left(x\right)=\int\frac{\mbox{d}\omega}{\pi}\frac{\Gamma_{L}\left(\omega\right)f_{L}\left(\omega\right)+\Gamma_{R}\left(\omega\right)f_{R}\left(\omega\right)}{\left(\omega-\varepsilon\left(x\right)\right)^{2}+\Gamma^{2}\left(\omega\right)}\label{eq:neff}
\end{eqnarray}
is the average, out-of-equilibrium, dot population valid for the adiabatic
limit~\cite{kosov_nonequilibrium_2009}. We find that the adiabatic
tunneling times for $\omega_{c}=25\mbox{cm\ensuremath{^{-1}}}$ and
$\omega_{c}=100\mbox{cm\ensuremath{^{-1}}}$ are of the order of $1500\frac{\hbar}{\Gamma}$
and $150\frac{\hbar}{\Gamma}$, respectively and the tunneling probabilities
are smaller than $10^{-5}$. For the former case ($\omega_{c}=25\mbox{cm\ensuremath{^{-1}}}$)
one may argue that this timescale is too long to be captured by the
RDM formalism with a cutoff time of $t_{c}\approx35\frac{\hbar}{\Gamma}$
and perhaps, for larger cutoff times which are not accessible to us,
the RDM will decay due to tunneling between the two wells. However,
this argument seems much less likely for $\omega_{c}=100\mbox{cm\ensuremath{^{-1}}}$,
where the tunneling time is much smaller ($150\frac{\hbar}{\Gamma}$),
and thus, tunneling should be captured even with cutoff times of the
order of $t_{c}\approx35\frac{\hbar}{\Gamma}$.

\begin{figure}[t]
\includegraphics[width=8cm]{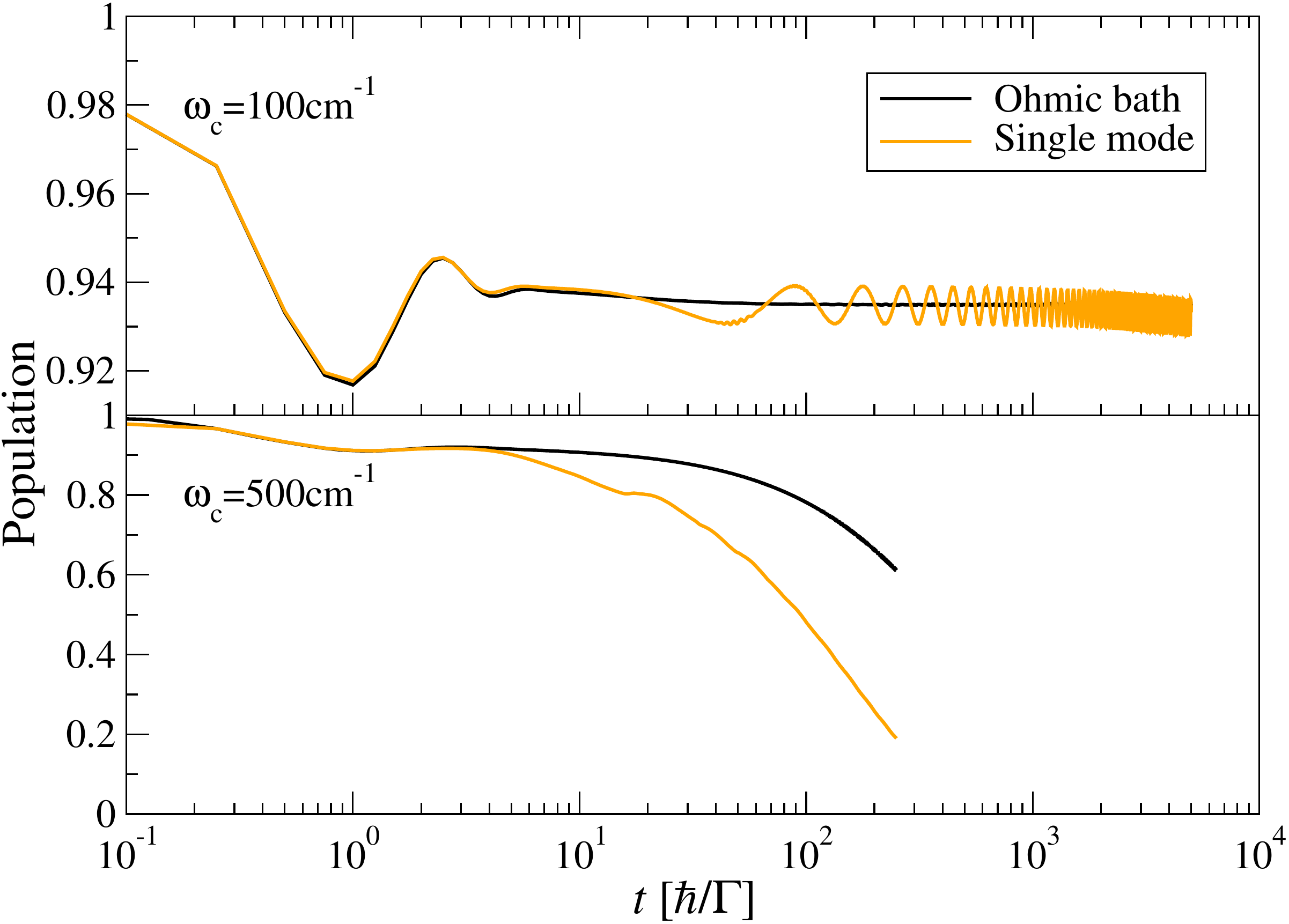} \caption{Dynamics of the dot population for an initially occupied bridge state
coupled to a single lead and a single phonon (orange) or an Ohmic
bath (black), for $\varepsilon_{d}/\Gamma=\frac{25}{8}$, $\lambda/\Gamma=2.7$,
and a bias of $\frac{5}{8}\Gamma$. $ $}

\label{fig:michael} 
\end{figure}

The fact that we do not observe any long time relaxation to a unique
steady state in the adiabatic limit is consistent with the notion
that tunneling is suppressed by the dynamical coupling to the phonons,
which was assumed static in the above estimation of the tunneling
process. Additionally, the low tunneling probability may also be used
to explain the vanishing long time transient behavior in the adiabatic
limit. To further elaborate on this and to elucidate the underlying
time scales and mechanisms, we have considered the simpler scenario
of the decay of an initially occupied dot state coupled only to the
unoccupied states in the right lead, \textit{i.e.} the states above
the chemical potential of the right electrode. This simplified version
of the Anderson-Newns model of heterogeneous electron transfer reduces
dramatically the computational complexity of the ML-MCTDH-SQR calculations
and allows us to directly access times that are of the order or longer
than the adiabatic tunneling times. In the upper panel of Fig.~\ref{fig:michael}
we show the population dynamics corresponding to this case for $\omega_{c}=100\mbox{cm\ensuremath{^{-1}}}$
and $\lambda/\Gamma=2.7$, for an initially occupied dot and shifted
phonon distribution ($\delta_{\alpha}=1$). We consider both a single
phonon mode and an Ohmic bath. The estimated adiabatic tunneling time
on $U_{{\rm eff}}(x)$ for this case is $300\frac{\hbar}{\Gamma}$.
The results for a single phonon mode show relaxation of the dot population
on timescales exceeding $10^{3}\frac{\hbar}{\Gamma}$, which indicate
that the dynamical coupling to a single mode increases the tunneling
time between the two wells compared with the pure adiabatic limit.
For the Ohmic bath, the dot population is stable even on times approaching
$10^{4}\frac{\hbar}{\Gamma}$ and tunneling is not observed, suggesting
stronger localization. This localization can be understood in term
of the reaction mode representation of the phonons, which for an Ohmic
bath corresponds to an over-damped oscillator.\cite{Thoss01,Wang2008,Weissbook}
Whether localization will suppress tunneling even at longer times
remains an open problem.

The lower panel of Fig.~\ref{fig:michael} shows the results for
the same simplified Anderson-Newns model, but for $\omega_{c}=500\mbox{cm\ensuremath{^{-1}\approx0.4\Gamma/\hbar}}$,
which is near the adiabatic/non-adiabatic crossover. The remaining
parameters are the same as those shown in the upper panel of Fig.~\ref{fig:michael}
for $\omega_{c}=100\mbox{cm\ensuremath{^{-1}\approx0.08\Gamma/\hbar}}$.
The dot population shows a two-step relaxation even for the Ohmic
case, eventually, relaxing to zero. The analysis shows that the longer
time decay can be associated with a non-adiabatic transition, with
a time constant that can be approximated by $\tau_{{\rm na}}\approx\frac{\hbar}{\Gamma}e^{\lambda/\hbar\omega_{c}}$
for the single-mode case.\cite{Schiller2013,Albrecht2013} Comparing
the single-mode to the Ohmic case reveals that the non-adiabatic transition
is much slower for the latter. This behavior is similar to the dynamics
of the population in the adiabatic limit, which showed vanishing tunneling
for the Ohmic case.

By analogy, we can associate the long-time decay of the full extended
Holstein model with two leads (right column-panels of Fig.~\ref{fig:long-time})
to a non-adiabatic transition from the occupied to the unoccupied
state. Despite the fact that the dot population does not decay to
zero, the timescales and behavior are similar to the single-lead case,
and the decay rate scales roughly as $e^{-\lambda/\hbar\omega_{c}}$.
Interestingly, the non-adiabatic transition does not destroy the bistability
(in some cases). This is rather surprising, but also very significant.
Despite having transitions between the two diabatic surfaces, the
long time limit plateau of the RDM still depends on the initial phonon
distribution! We note in passing that the NEGF-SCBA approach does
not describe the non-adiabatic process (dashed curves in Fig.~\ref{fig:long-time}),
and therefore, does not show any long time transient behavior in this
parameter regime.

\section{Signatures of Bistability\label{sec:bistability}}

\begin{figure}[t]
\includegraphics[width=9cm]{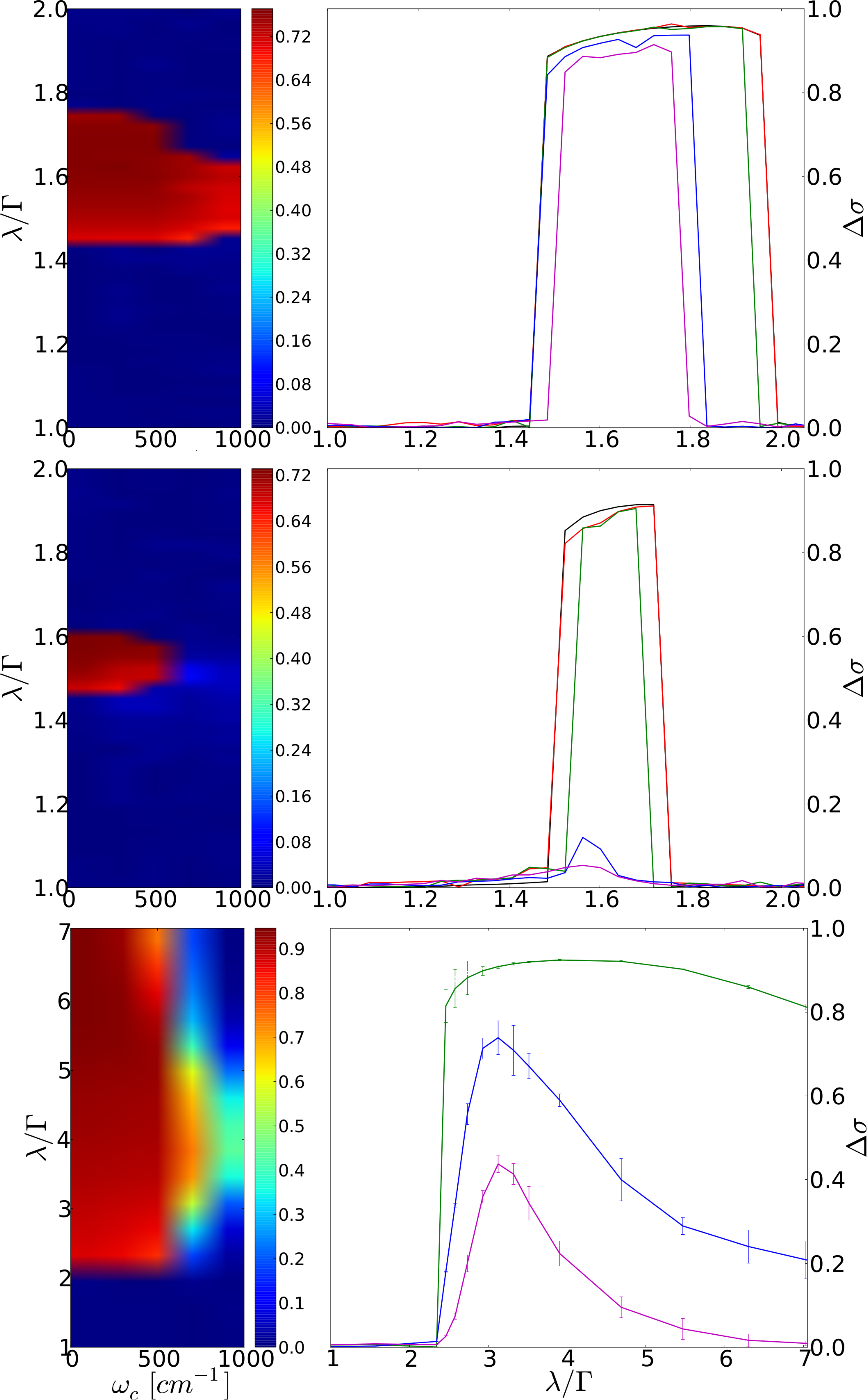}

\caption{A plot of the value of the bistability, $\Delta\sigma$, as a function
of $\lambda$ and $\omega_{c}$. Upper, middle and lower panels are
for $\varepsilon_{d}/\Gamma=\frac{25}{16}$ and $\mu_{L}-\mu_{R}=0$,
$\varepsilon_{d}/\Gamma=\frac{25}{16}$ and $\mu_{L}-\mu_{R}=\frac{5}{8}\Gamma$,.
and $\varepsilon_{d}/\Gamma=\frac{25}{8}$ and $\mu_{L}-\mu_{R}=\frac{5}{8}\Gamma$,
respectively. The upper two panels were generated by the steady-state
NEGF-SCBA approach. For the lower panel, the steady-state NEGF-SCBA
was used for $\lambda/\Gamma\leq2\frac{1}{3}$ and ML-MCTDH-SQR combined
with the RDM otherwise. Black, red, dark green, blue and magenta show
results for $\omega_{c}=25,50,100,500$ and $1000\mbox{cm\ensuremath{^{-1}}(\ensuremath{\approx0.02}, \ensuremath{\approx0.04}, \ensuremath{\approx0.08}, \ensuremath{\approx0.4}and \ensuremath{\approx0.8}in units of \ensuremath{\Gamma/\hbar})}$,
respectively.}

\label{fig:diagram} 
\end{figure}

We have shown previously that the value of the RDM at steady state
is independent of the initial occupation of the dot, i.e.\ on the
initial state of the electronic degrees of freedom.\cite{wilner_bistability_2013}
The proof is rather simple and is based on the Laplace final value
theorem which relates $\sigma\left(t\rightarrow\infty\right)$ to
the integral of the memory kernel, $\mathcal{K}=\frac{1}{\hbar^{2}}\int_{0}^{\infty}d\tau\kappa\left(\tau\right)$.
Indeed, for all the results shown above, the long time limit of the
RDM is independent of the initial dot occupation, as it should be.
However, for certain model parameters, we find (and also others~\cite{galperin_hysteresis_2005,galperin_non-linear_2008,Kosov2011,Kosov2011a,albrecht_bistability_2012,wilner_bistability_2013})
that the long time value of the RDM can depend on the initial preparation
of the phonon degrees of freedom. This finding suggests the existence
of bistability in the system. The value of the population difference
of the two initial phonon preparation, $\Delta\sigma=\sigma_{11}^{\delta_{\alpha}=1}\left(t\rightarrow\infty\right)-\sigma_{11}^{\delta_{\alpha}=0}\left(t\rightarrow\infty\right)$
for long times is a measure of the importance of bistability, and
will in the following be referred to simply as bistability. In the
current section, we analyze the dependence of bistability on the various
model parameters.

In Fig.~\ref{fig:diagram} we plot the results for $\Delta\sigma$
for two values of the dot energy $\varepsilon_{d}$ and the bias voltage
$\Delta\mu=\mu_{L}-\mu_{R}$, and for a range of frequencies ($\omega_{c}$)
and reorganization energies ($\lambda$). The upper two panels correspond
to $\varepsilon_{d}/\Gamma=25/16$. In this case, the results were
generated using the steady-state NEGF-SCBA and thus, the approach
is limited to relatively low values of $\lambda$. Note, however,
that bistability is not observed for $\lambda/\Gamma>2$, which is
exactly the regime where NEGF-SCBA is accurate, as shown above (cf.,
Fig.~\ref{fig:comparison}). In fact, comparing the dynamics for
one of the values of $\omega_{c}$ generated by the NEGF-SCBA with
the numerically converged ML-MCTDH-SQR for which $\Delta\sigma\ne0$
indicates excellent agreement (data not shown here) even for $\varepsilon_{d}/\Gamma=25/16$.

\begin{figure}[H]
\centering{}\includegraphics[width=6cm]{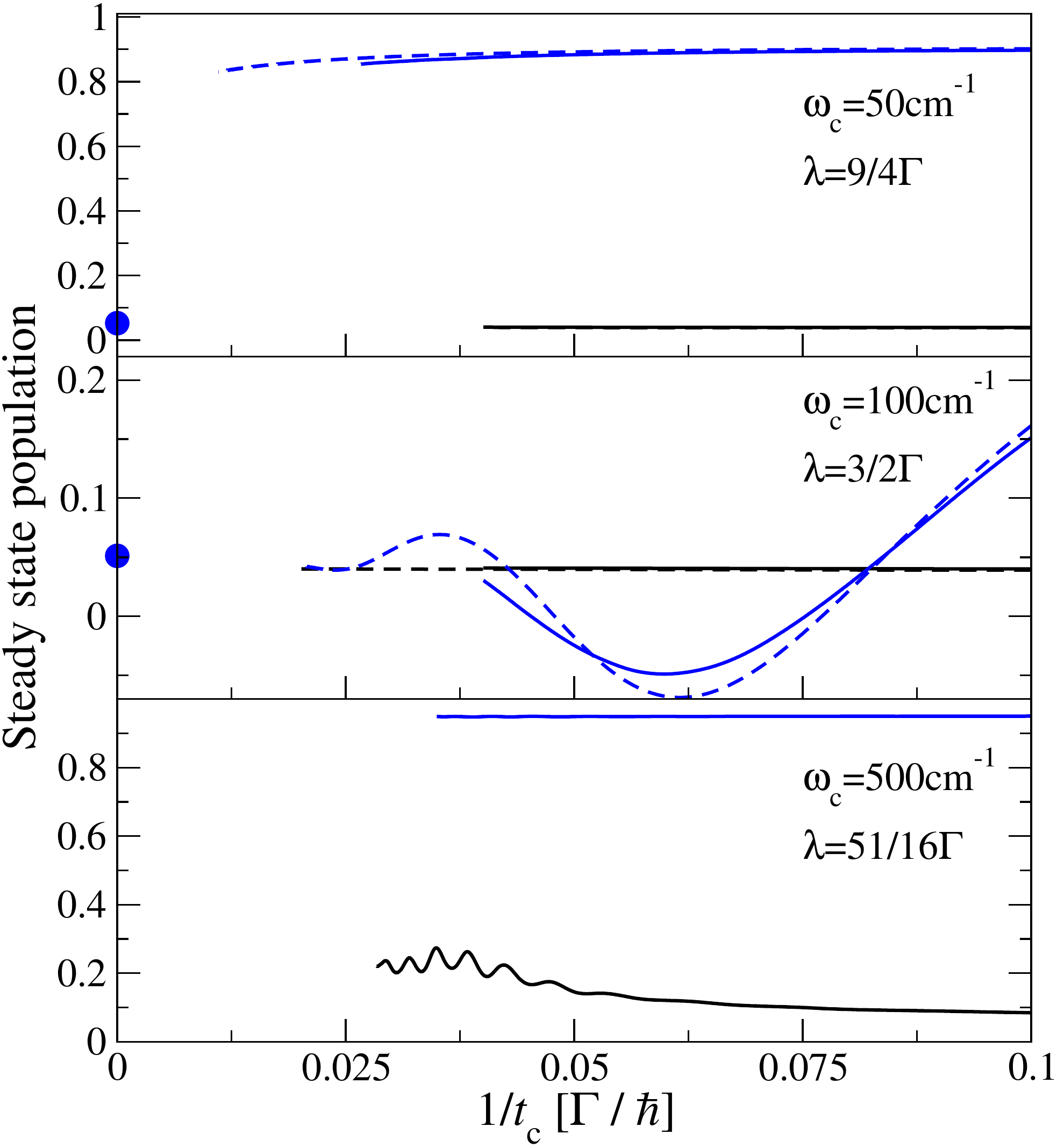}\caption{The steady-state dot population as a function of $1/t_{c}$. Solid
and dashed curves are results obtain by the ML-MCTDH-SQR and NEGF-SCBA
combined with the RDM formalism, respectively. The black and blue
lines are results for $\delta_{\alpha}=0$ and $1$, respectively.
The solid circle is the steady-state NEGF-SCBA result. $\varepsilon_{d}/\Gamma=\frac{25}{8}$
for all panels.}

\label{fig:1/tc} 
\end{figure}

In the lower panel of Fig.~\ref{fig:diagram} we show results for
$\varepsilon_{d}/\Gamma=25/8$ and $\Delta\mu=\frac{5}{8}\Gamma$.
Here, the results were generated by the ML-MCTDH-SQR approach combined
with the RDM formalism and thus, are not limited to small values of
$\lambda$. In most cases, we used a cutoff time $t_{c}\le35\frac{\hbar}{\Gamma}$.
This cutoff time was not always sufficient to converge the long time
values of the RDM. In the upper panel of Fig.~\ref{fig:1/tc} we
illustrate this for a case where $\Delta\sigma=0$ and for a sufficiently
small value of $\lambda$ so that the ML-MCTDH-SQR results can be
compared with the NEGF-SCBA. For the initial condition corresponding
to $\delta_{\alpha}=0$ (black curves), we find that the values of
the RDM are insensitive to the cutoff time for $t_{c}\ge10\frac{\hbar}{\Gamma}$.
This is expected, since the steady state of the system is close to
the initial condition $\delta_{\alpha}=0$ and thus, the phonons are
nearly at steady state initially. This is not the case for the other
initial condition corresponding to $\delta_{\alpha}=1$ (blue curves).
As $\frac{1}{t_{c}}$ decreases the dot population decreases and never
levels off. In fact, the steady-state value of the dot population
obtained from the steady-state NEGF-SCBA is rather small and equals
that value for $\delta_{\alpha}=0$ (\textit{i.e.}, $\Delta\sigma=0$).
In this case, it seems that a much larger cutoff time is needed to
converge the RDM in this case, even larger than the limit of the two-time
NEGF-SCBA which is $t_{c}\approx100\frac{\hbar}{\Gamma}$.

The middle panel of Fig.~\ref{fig:1/tc} shows results for a relatively
small coupling parameter for which the ML-MCTDH-SQR results can be
compared with those of the NEGF-SCBA approach. Again, for $\delta_{\alpha}=0$,
a rather small cutoff time is sufficient to converge the results since
the phonon initial density matrix is close to its steady-state value.
The case of $\delta_{\alpha}=1$ requires a much larger cutoff time.
In fact, larger than the computational limit of the ML-MCTDH-SQR approach,
but still within the reach of the two-time NEGF-SCBA, for which a
clear plateau is observed as $\frac{1}{t_{c}}$ decreases. The plateau
value agrees well with the steady-state NEGF-SCBA calculation (solid
circle). Situations of this sort are consider converged.

In the lower panel of Fig.~\ref{fig:1/tc} we show results for a
large value of $\lambda/\Gamma>3$, and thus, only the ML-MCTDH-SQR
was used to obtain the RDM. Here, the well corresponding to $\delta_{\alpha}=1$
is the more stable one, and therefore, it is rather easy to converge
the dot population for this initial condition (blue curve). For the
other initial condition, a clear leveling of the dot population as
$\frac{1}{t_{c}}\rightarrow0$ is evident. However, the value of the
steady-state is quite noisy due to computational limitations of the
ML-MCTDH-SQR method. Situations of this sort, for which we observe
the beginning of the leveling of the dot population as $t_{c}$ is
increased to the computational limit, will be considered converged.
However, to indicate the fact that the long time limit of the dot
population is noisy, we assign a large error bar of the size of the
fluctuations to the value of $\Delta\sigma$ shown in Fig.~\ref{fig:diagram}.

Returning to discuss the results of Fig.~\ref{fig:diagram} within
the above limitations concerning the convergence of the results, several
important conclusion can be drawn: 
\begin{itemize}
\item As the source-drain bias voltage $V$ increases the window of bistability
decreases and will eventually disappear.\cite{galperin_hysteresis_2005}
It is important to note, however, that we find a finite value for
$\Delta\sigma$ on timescales much longer than $\frac{\hbar}{\Delta\mu}$.
A similar effect is expected if the temperature is increased. 
\item The window of bistability also decreases as the dot energy $\varepsilon_{d}$
decreases. For the adiabatic limit, this is strongly correlated with
the range of reorganization energies, $\lambda$, for which the effective
potential of the phonons shows a distinct double-well structure. This
range decreases with $\varepsilon_{d}$. 
\item As $\omega_{c}$ increases the window of bistability decreases and
so does the value of $\Delta\sigma$. Surprisingly, however, even
for relatively large values of $\hbar\omega_{c}\approx\Gamma$ away
from the adiabatic limit, we still observe bistability. 
\end{itemize}
In the adiabatic limit, the first two findings can be rationalized
by the already mentioned fact that a precondition for bistability
is the existence of an effective potential for the phonons with two
stable minima, which have to have energies outside the bias window,
\textit{i.e.} $\varepsilon_{d}-2\lambda\ll\mu_{L/R}\ll\varepsilon_{d}$
and $\Gamma,V\ll\lambda$ (see also the discussion in Refs.\ \onlinecite{galperin_hysteresis_2005}
and \onlinecite{Mozyrsky05}). The most striking result is that the
phenomenon of bistability exists away from the strictly adiabatic
limit and prevails on time scales longer than the non-adiabatic transition
time, i.e.\ on much longer time scales than previously thought.\cite{albrecht_bistability_2012}
The question remains, however, whether bistability in the extended
Holstein model exists in the strict long-time limit. The unambiguous
clarification of this question requires a numerically exact methodology
which can address directly the long-time limit of this model, which
is yet to be developed.

\section{Concluding Remarks\label{sec:Conclusions}}

In this paper, we have investigated the nonequilibrium quantum dynamics
of the extended Holstein model as a generic model for charge transport
in a quantum dot with electron-phonon interactions. We have specifically
focused on the transient dynamics and the approach to steady-state.
To this end, we have used a methodology, which combines a reduced
density matrix formalism based on projection-operator techniques and
two different approaches to calculate the memory kernel, a two-time
NEGF with the SCBA and the ML-MCTDH-SQR. The latter method provides
a numerically exact treatment of the many-body quantum dynamics up
to a certain time.

The results obtained in a wide range of parameters reveal dynamics
on multiple timescales. In addition to the short and intermediate
timescales associated with the separate electronic and phononic degrees
of freedom, the electron-phonon coupling introduces longer timescales
related to the adiabatic or nonadiabatic tunneling between the two
charge states. The analysis shows, furthermore, that the value of
the dot occupation may depend on the initial preparation of the phonon
degrees of freedom, suggesting the existence of bistability. Intriguingly,
the phenomenon of bistability persists even on timescales longer than
the adiabatic/nonadiabatic tunneling time. Considering different parameter
ranges, we have formulated conditions for bistability. This analysis
shows that bistability is particularly pronounced for low characteristic
frequencies of the phonons and moderate to large electron-phonon couplings.
On the other hand, bistability is quenched for larger voltages. A
similar effect is expected for higher temperatures.

The present study, employing time-dependent methods, cannot address
the strict long-time limit and, therefore, cannot give a final answer
to the controversial question whether a unique steady-state always
exists for the extended Holstein model. The results do show, however,
a significant dependence on the initial state on timescales which
are accessible by time-resolved spectroscopy and, thus, should be
experimentally observable.

\section{Acknowledgments}

EYW and ER would like to thank Tal Levy, Kristen Kaasbjerg and Robert
van Leeuwen for insightful discussions. MT thanks Jeremy Richardson
for helpful discussions. EYW is grateful to The Center for Nanoscience
and Nanotechnology at Tel Aviv University for a doctoral fellowship.
HW acknowledges the support from the National Science Foundation CHE-1012479.
This work was supported by the German Research Council (DFG) and used
resources of the National Energy Research Scientific Computing Center,
which is supported by the Office of Science of the U.S. Department
of Energy under Contract No. DE-AC02-05CH11231, and the Leibniz Computing
Center (LRZ) Munich.

\end{document}